\documentclass[aps,prd,twocolumn,superscriptaddress]{revtex4-2}

\usepackage[utf8]{inputenc}
\usepackage{amsmath}
\usepackage{physics}
\usepackage{comment}
\usepackage{bm}
\usepackage[separate-uncertainty=true]{siunitx}
\usepackage{graphicx}
\usepackage{natbib}
\usepackage[hidelinks]{hyperref}
\usepackage{mathtools}
\usepackage{xcolor}
\usepackage{ctable}
\usepackage[normalem]{ulem}
\usepackage{multirow}
\usepackage[caption=false]{subfig}
\usepackage[export]{adjustbox}

\newcommand{\rmd}{{\rm d}}
\newcommand{\jac}[2]{\abs{\frac{\partial #1}{\partial #2}}}
\newcommand{\chieff}{{\chi_{\rm eff}}}
\newcommand{\cumchidiff}{{C_{\rm diff}}}
\newcommand{\mchirp}{{\mathcal{M}}}
\newcommand{\thetanet}{{\theta_{\rm net}}}
\newcommand{\phinet}{{\phi_{\rm net}}}
\newcommand{\phinethat}{{\hat\phi_{\rm net}}}
\newcommand{\phiref}{\phi_{\rm ref}}
\newcommand{\trefdet}{t_{k_0}}
\newcommand{\dhat}{\hat d}
\newcommand{\dl}{d_L}
\newcommand{\sonex}{\chi_{1x}}
\newcommand{\soney}{\chi_{1y}}
\newcommand{\sonez}{\chi_{1z}}
\newcommand{\stwox}{\chi_{2x}}
\newcommand{\stwoy}{\chi_{2y}}
\newcommand{\stwoz}{\chi_{2z}}
\newcommand{\sonexn}{\chi_{1x}^{N}}
\newcommand{\soneyn}{\chi_{1y}^{N}}
\newcommand{\stwoxn}{\chi_{2x}^{N}}
\newcommand{\stwoyn}{\chi_{2y}^{N}}
\newcommand{\soner}{\chi_1^\perp}
\newcommand{\stwor}{\chi_2^\perp}
\newcommand{\csoner}{C_1^\perp}
\newcommand{\cstwor}{C_2^\perp}
\newcommand{\gmst}{{\rm GMST}} 
\newcommand{\fbar}{\overline{f}}
\newcommand{\tc}{t_c}
\newcommand{\los}{\bm{\hat n}}
\newcommand{\propagation}{\bm{\hat N}}
\newcommand{\phirefhat}{\hat\phi_{\rm ref}}
\newcommand{\phijlhat}{\hat\phi_{JL}}
\newcommand{\thetajn}{\theta_{JN}}
\newcommand{\fref}{{f_{\rm ref}}}
\newcommand{\tabline}{\specialrule{.1em}{.05em}{.05em}}

\begin{document}

\title{
Removing degeneracy and multimodality in gravitational wave source parameters
}

\author{Javier Roulet}
\email{jroulet@ucsb.edu}
\affiliation{\mbox{Kavli Institute for Theoretical Physics, University of California at Santa Barbara, Santa Barbara, CA 93106, USA}}

\author{Seth Olsen}
\affiliation{\mbox{Department of Physics, Princeton University, Princeton, NJ 08540, USA}}

\author{Jonathan Mushkin}
\affiliation{\mbox{Department of Particle Physics \& Astrophysics, Weizmann Institute of Science, Rehovot 76100, Israel}}

\author{Tousif Islam}
\affiliation{\mbox{Department of Physics, University of Massachusetts, Dartmouth, MA 02747, USA}}
\affiliation{\mbox{Department of Mathematics, University of Massachusetts, Dartmouth, MA 02747, USA}}
\affiliation{\mbox{Center for Scientific Computing and Visualization Research, University of Massachusetts, Dartmouth, MA 02747, USA}}
\affiliation{\mbox{Kavli Institute for Theoretical Physics, University of California at Santa Barbara, Santa Barbara, CA 93106, USA}}

\author{Tejaswi Venumadhav}
\affiliation{\mbox{Department of Physics, University of California at Santa Barbara, Santa Barbara, California 93106, USA}}
\affiliation{\mbox{International Centre for Theoretical Sciences, Tata Institute of Fundamental Research, Bangalore 560089, India}}

\author{Barak Zackay}
\affiliation{\mbox{Department of Particle Physics \& Astrophysics, Weizmann Institute of Science, Rehovot 76100, Israel}}

\author{Matias Zaldarriaga}
\affiliation{\mbox{School of Natural Sciences, Institute for Advanced Study, 1 Einstein Drive, Princeton, New Jersey 08540, USA}}

\date{\today}
             
\begin{abstract}
Quasicircular binary black hole mergers are described by 15 parameters, of which gravitational wave observations can typically constrain only $\sim 10$ independent combinations to varying degree.
In this work, we devise coordinates that remove correlations, and disentangle well- and poorly-measured quantities.
Additionally, we identify approximate discrete symmetries in the posterior as the primary cause of multimodality, and design a method to tackle this type of multimodality.
The resulting posteriors have little structure and can be sampled efficiently and robustly.
We provide a Python package for parameter estimation, \texttt{cogwheel}, that implements these methods together with other algorithms for accelerating the inference process.
One of the coordinates we introduce is a spin azimuth that is measured remarkably well in several events.
We suggest this might be a sensitive indicator of orbital precession, and we anticipate that it will shed light on the occurrence of spin--orbit misalignment in nature.
\end{abstract}

\maketitle

\section{Introduction}

Gravitational wave astronomy is advancing at a rapid pace, with over 100 signals from compact binary mergers detected since the advanced LIGO \cite{Aasi2015} and advanced Virgo \cite{Acernese2014} detectors became operational \cite{Abbot2019_GWTC-1, Nitz2019_1-OGC, Zackay2019_GW151216, Venumadhav2020, Nitz2020_2-OGC, Zackay2021, Abbott2021_GWTC-2, Abbott2021_GWTC-2.1, Abbott2021_GWTC-3, Nitz2021_3-OGC, Olsen2022}.
The rate of detections will keep accelerating as planned and ongoing hardware upgrades take place.
The astrophysical interpretation of these detections requires measuring their source parameters, such as masses, spins, position and orientation.

At the same time, following recent developments in waveform modeling, state-of-the-art models of compact binary coalescences now incorporate higher-order harmonics and generically oriented spins \cite{Varma2019, Ossokine2020, Khan2020, Pratten2021}.
These improvements have increased the diversity of waveform morphologies and the parameter space dimensionality, rendering computational cost a major hurdle for parameter estimation studies.

In this work we develop a new parametrization of the source properties, tailor-made for compact binary mergers, that removes most correlations and multimodalities encountered in their posterior distributions.
This coordinate system emphasizes quantities that are best constrained by the data, as opposed to the physically motivated parameters used by external libraries to model waveforms.
We derive the set of analytic, invertible, nonlinear transformations between the standard parameter space and this new coordinate system.
Expressed in these coordinates, the posterior distributions exhibit much less structure, which makes it easier to sample them efficiently and robustly.

In general, under a coordinate change $x \to y$ the prior and posterior probability densities transform as
\begin{equation}
    P(y) = P(x)\abs{J},
\end{equation}
where
\begin{equation}
    \abs{J} = \abs{\partial x / \partial y} 
\end{equation}
is the the absolute value of the Jacobian determinant.
This can make coordinate changes cumbersome if their Jacobian determinant is not easily computable.
We define our coordinate transformation as a series of simple transformations, affecting few variables at a time, that have tractable Jacobian determinants.
We will use the following properties of the Jacobian.
First, they multiply under composition of transformations $x \to y \to z$:
\begin{equation}
    \jac{x}{z} = \jac{x}{y} \cdot \jac{y}{z}.
\end{equation}
Second, transformations of the form
\begin{equation}\label{eq:shear}
    \begin{pmatrix}
        x_1 \\
        x_2
    \end{pmatrix}
    \to
    \begin{pmatrix}
        a(x_2) x_1 + b(x_2) \\
        x_2
    \end{pmatrix}
\end{equation}
(where $x_1, x_2$ may be multidimensional) have
\begin{equation}
    \abs{J} = \abs{a(x_2)},
    \label{eq:jac}
\end{equation}
which frees us to use arbitrarily sophisticated $b(x_2)$ functions.
In particular, if $a(x_2) = \pm 1$ the probability density is unchanged by the coordinate transformation in Eq.~\eqref{eq:shear}.

The remainder of the article is structured as follows.
In Sec.~\ref{sec:likelihood} we review the likelihood function of gravitational wave source parameters.
In Sec.~\ref{sec:coordinates} we introduce coordinates that remove common correlations.
In Sec.~\ref{sec:multimodality} we identify approximate symmetries responsible for discrete degeneracies, and present a novel method to remove this type of multimodality.
In Sec.~\ref{sec:results} we present our main results: Sec.~\ref{ssec:performance} assesses the performance improvement brought by our methods in terms of the accuracy of the recovered posterior and the computational cost, and Sec.~\ref{ssec:azimuth} shows that a particular spin azimuth can be measured surprisingly well.
We conclude in Sec.~\ref{sec:conclusions} with a summary and outlook.
Appendix~\ref{app:symmetries} discusses in further detail two of the approximate symmetries identified, and appendix~\ref{app:reference} acts as a reference sheet in which we summarize the coordinate system proposed.

\section{Likelihood}
\label{sec:likelihood}

In this section we briefly review the likelihood function $P(d \mid \theta)$, through which the data $d$ constrain source parameters $\theta$.
We cast its approximate dependence on extrinsic parameters in terms of the signal's amplitude, phase and time of arrival at each detector, and we provide analytical expressions for these.

Under the approximation that the noise is stationary and Gaussian,
\begin{equation}
    \log P(d \mid \theta)
        = - \frac 12 \sum_{k \in {\rm det}} 4\int_0^\infty \rmd f \frac{\abs{\tilde d_k(f) - \tilde h_k(f; \theta)}^2}{S_k(f)},
\end{equation}
where $\tilde d_k$, $\tilde h_k$ and $S_k$ are the frequency-domain data, strain model, and one-sided noise power spectrum in the $k$th detector, respectively~\cite{Jaranowski2005}.

For quasicircular binaries, the detector strains $\tilde h_k$ depend on 15 parameters: eight intrinsic (two masses $m_1, m_2$ and six components of the spin vectors $\bm \chi_1, \bm \chi_2$) and seven extrinsic (luminosity distance $\dl$, inclination $\iota$, right ascension $\alpha$, declination $\delta$, polarization $\psi$, orbital phase $\phiref$ and coalescence time $t_c$).
We can understand the approximate dependence on these parameters by considering the waveform under the quadrupole, stationary-phase and post-Newtonian approximations:
\begin{multline}
    \tilde h_k(f; \theta) \\
    \approx
        A(f)
        \frac{\mchirp^{5/6}}{\dl}
        R_k(\iota, \los, \psi)
        e^{i [2 \phiref
        - 2\pi f t_k(\los)
        + \Psi(f; \theta_{\rm int})]}, \label{eq:h22np}
\end{multline}
where the chirp mass
\begin{equation}
    \mchirp = \frac{(m_1 m_2)^{3/5}}{(m_1 + m_2)^{1/5}}
\end{equation}
controls the waveform amplitude and phase to leading post-Newtonian order.
The inclination, sky location and polarization enter through the detector response \cite{Cutler1994}
\begin{equation}
\label{eq:response}
  \begin{split}
    R_k &= \frac{1 + \cos^2\iota}{2} F^+_k (\los, \psi)
          -i \cos\iota\, F^\times_k (\los, \psi);\\
  \end{split}
\end{equation}
where $\los$ is the line of sight and $F^+_k, F^\times_k$ are the antenna factors of the $k$th detector \cite{Whelan2013}.
The sky location additionally introduces an arrival time delay in each detector due to the finite speed $c$ of gravitational waves, so there is a separate arrival time $t_k$ at each detector $k$.

Altogether, to this level of approximation the extrinsic parameters only affect the amplitude, phase and time in each detector:
\begin{equation}
    \tilde h_k(f; \theta)
    \approx
        a_k
        e^{i \varphi_k}
        e^{-i2\pi (f-\overline f_k) t_k}
        A(f) e^{i\Psi(f; \theta_{\text{int}})},
\end{equation}
with
\begin{align}
    a_k &= \frac{\mchirp^{5/6}}{\dl} \abs{R_k(\iota, \los, \psi)}
        \label{eq:a_k}\\
    \varphi_k &= \arg{R_k(\iota, \los, \psi)} + 2 \phiref
        - 2\pi \fbar_k t_k \label{eq:varphi_k}\\
    t_k &= t_c - \frac{\los \cdot \bm r_k}{c}, \label{eq:t_k}
\end{align}
where $\bm r_k$ is the detector location relative to the point of reference time, e.g.\ the center of Earth, and $\fbar_k$ is a frequency scale defined below.
In Eq.~\eqref{eq:varphi_k} we have included a term dependent on $t_k$ to make $\varphi_k$ orthogonal to $t_k$: to linear order, it is possible to absorb a small shift in the time of arrival $t_k$ (in the sense of leaving the observed waveform almost unchanged) by simultaneously changing the phase (via $\phiref$, say) by $2 \pi \overline f_k t_k$  \cite{Fairhurst2009,Roulet2019}, where $\overline f_k$ is the first frequency moment, i.e.,
\begin{equation}
    \overline {f^n}_k \coloneqq \frac{\int_0^\infty \rmd f A^2(f) f^n / S_k(f)}{\int_0^\infty \rmd f A^2(f) / S_k(f)} \label{eq:fmoment}
\end{equation}
for $n=1$.
This way, we expect the measurements of $a_k, \varphi_k, t_k$ in a detector to be uncorrelated.
We can also estimate the uncertainties in each of these parameters, which scale inversely with the signal-to-noise ratio $\rho_k$:
\begin{align}
    \Delta a_k &\sim \frac {a_k}{\rho_k} \label{eq:da}\\
    \Delta \varphi_k &\sim \frac {1}{\rho_k} \label{eq:dphi}\\
    \Delta t_k &\sim \frac {1}{2 \pi \sigma_f \rho_k} \label{eq:dt},
\end{align}
with $\sigma_f^2 = \overline{f^2} - \overline f^2$ \cite{Cutler1994, Fairhurst2009}.
By convention, for each event we will sort the detectors by their signal-to-noise ratio ($\rho_{k_0} > \rho_{k_1} > \ldots$), so the best measured quantities correspond to the ``reference detector" $k_0$.

\section{Mitigating degeneracy}
\label{sec:coordinates}

Oriented by the previous discussion, in this section we introduce a system of coordinates in which some variables separately control the observables $a_{k_0}, \varphi_{k_0}, t_{k_0}, t_{k_1}-t_{k_0}$, which will typically be well measured, and others explore degeneracies and will be poorly constrained.
Some of these coordinates are already known in the literature but we still review them here for completeness.
As we will see, the posterior distribution expressed in these coordinates is largely uncorrelated.

\subsection{Amplitude at reference detector}
\label{ssec:amplitude}

\begin{figure}
    \centering
    \includegraphics[width=\linewidth]{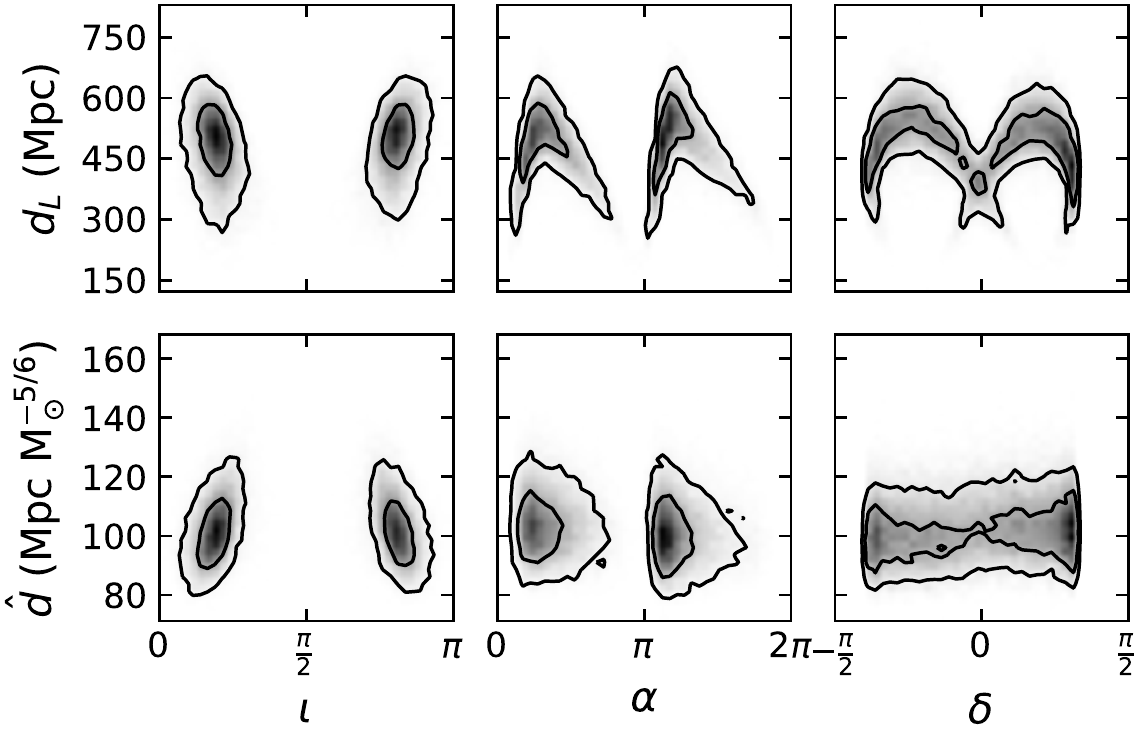}
    \caption{The luminosity distance $\dl$ is typically correlated with the inclination and sky location (top row), while $\dhat$ (Eq.~\eqref{eq:dhat}) is less correlated and better measured (bottom row).
    Examples in this section correspond to GW151226 \cite{Abbott2016_GW151226}, but the highlighted qualitative features are generic.}
    \label{fig:dhat}
\end{figure}

We replace the luminosity distance by the so-called chirp distance \cite{Brady2008}, which controls the amplitude at the reference detector Eq.~\eqref{eq:a_k}:
\begin{equation}
    \label{eq:dhat}
    \begin{split}
        \dhat(\dl, \mchirp, \iota, \alpha, \delta, \psi)
        &\coloneqq \frac{1}{a_{k_0}} \\
        &= \frac{\dl}{\mchirp^{5/6}\abs{R_{k_0}}}.
    \end{split}
\end{equation}
This is proportional to the luminosity distance, with a scale factor that depends on the response of the detector to the given $\iota, \alpha, \delta, \psi$, and the intrinsic amplitude of the source $\propto\mchirp^{5/6}$.
Thus, $\dhat$ avoids the correlations with these variables that the luminosity distance suffers, as shown in Fig.~\ref{fig:dhat}.
A collateral benefit is that the observable values of $\dhat$ are similar for all mass ranges (in contrast, heavier events can be observed at larger luminosity distances), so the distance range to explore does not need to be tuned for each event.
From Eqs.~\eqref{eq:da} and \eqref{eq:dhat} we expect that $\dhat$ will typically be measured to a precision $\Delta\dhat \sim \dhat / \rho_{k_0}$.

In the notation of Eq.~\eqref{eq:shear}, the transformation $\dhat \to \dl$ has $x_1 = \dhat$, $x_2=(\mchirp, \iota, \alpha, \delta, \psi)$, $a(x_2) = 1/{\mchirp^{5/6}\abs{R_{k_0}}}$ and $b = 0$.
The Jacobian determinant is
\begin{equation}
    \label{eq:dhat_jacobian}
    \abs{J} = \abs{\frac{\partial \dhat}{\partial \dl}}
    = \frac{1}{\mchirp^{5/6}\abs{R_{k_0}}}
    = \frac{\dhat}{\dl}.
\end{equation}
A more elaborate solution is to marginalize the posterior semianalytically over distance, altogether removing the necessity to sample from it \cite{Singer2016, RomeroShaw2020}.
We include this functionality as well in the software package we are releasing along with this paper.
In practice, this makes the sampling process more robust against a particular failure mode, in which the sampler occasionally explores the distant universe (favored by the prior) and misses a nearby solution with high likelihood.

\subsection{Time of arrival at reference detector}

We use the time of arrival at the reference detector $\trefdet$ as our arrival time parameter \cite{RomeroShaw2020}, as opposed to, for example, the geocentric time of arrival.

\begin{figure}
    \centering
    \includegraphics[width=\linewidth]{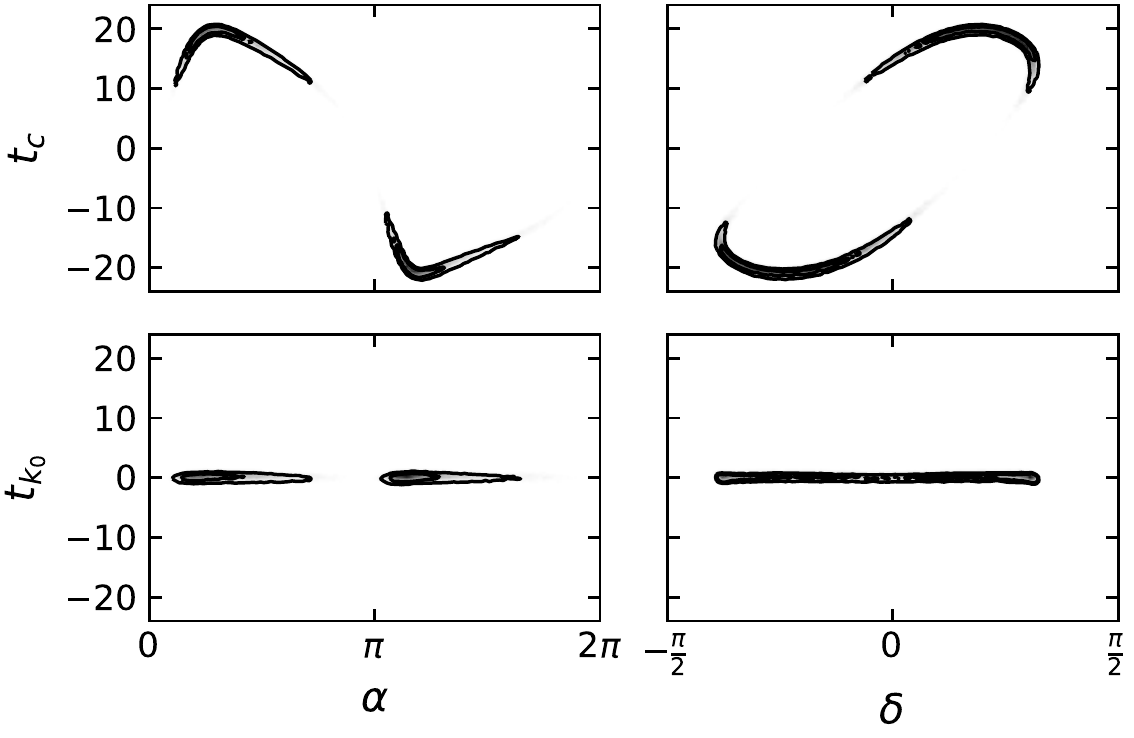}
    \caption{The arrival time at geocenter (top row) is worse measured than at the reference detector (bottom row), because the time delay introduces a large correlation with sky location.}
    \label{fig:tk0}
\end{figure}

From Eq.~\eqref{eq:dt}, the uncertainty in arrival time at the detector is typically $\lesssim \SI{1}{\milli\second}$ \cite{Fairhurst2009}.
On the other hand, unless the sky location is particularly well constrained, the uncertainty in time of arrival at geocenter is on the order of the gravitational wave Earth-crossing time (roughly $\SI{40}{\milli\second}$) due to the tight nonlinear correlation with sky location shown in Fig.~\ref{fig:tk0} for the case of GW151226.

From Eqs.~\eqref{eq:shear}, \eqref{eq:jac} and \eqref{eq:t_k}, the transformation $\trefdet \to \tc$ has unit Jacobian determinant.

\subsection{Time-of-arrival difference}
\label{ssec:time_difference}

When a signal is observed in multiple detectors, the arrival time differences provide the dominant constraint on its sky location.
For each pair of detectors, the arrival time difference determines the angle $\theta_{kk'}$ between the source location and an axis through both detectors according to (see Eq.~\eqref{eq:t_k})
\begin{equation}
\label{eq:dt_detectors}
    \begin{split}
        t_k - t_{k'}
        &= \los \cdot \frac{\bm r_{k'} - \bm r_k}{c} \\
        &= \tau_{k k'} \cos \theta_{k k'},
    \end{split}
\end{equation}
where $\tau_{kk'}$ is the the gravitational wave travel time between detectors $k$ and $k'$.

\begin{figure}
    \centering
    \includegraphics[width=.9\linewidth]{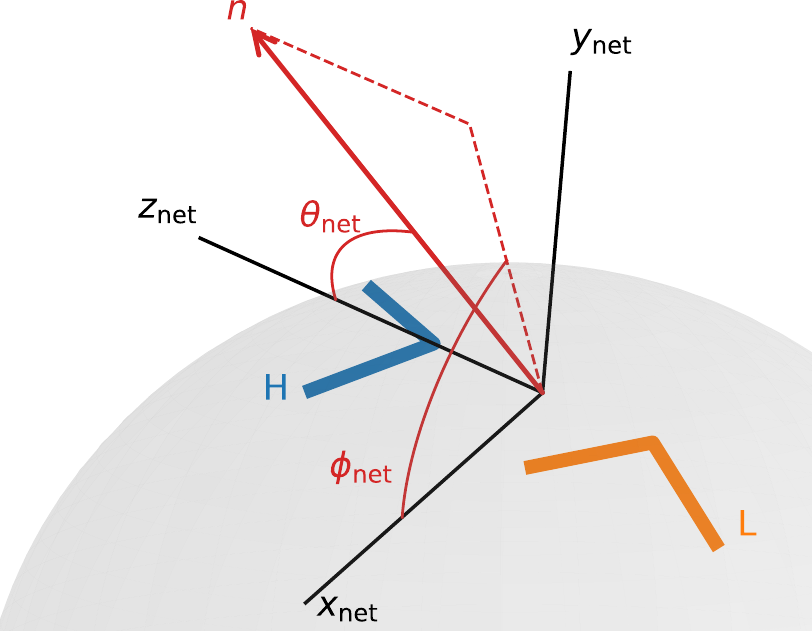}
    \caption{Coordinate system based on the dominant pair of detectors in the network, in this example Hanford--Livingston.
        The $z$-axis contains the two detectors and the $y$-axis points perpendicularly upwards.
        The zenithal angle $\thetanet$ of the line of sight $\los$ determines the time-of-arrival difference.
        (Detector sizes are exaggerated but otherwise the figure is to scale.)}
    \label{fig:ligo_angles}
\end{figure}

Thus, a natural way of parametrizing the sky location measurement is with a polar coordinate system $(\thetanet, \phinet)$ whose $z$-axis contains the two detectors with the largest signal-to-noise ratios in the network \cite{Veitch2010,RomeroShaw2020}, shown in Fig.~\ref{fig:ligo_angles}.
This coordinate system rotates with Earth and is related to the (fixed) $\alpha, \delta$ by a 3D rotation that depends on the pair of detectors and the sidereal time at which the signal arrives.
The isotropic prior is uniform in $\cos \thetanet, \phinet$.

\begin{figure}
    \centering
    \includegraphics[width=.97\linewidth]{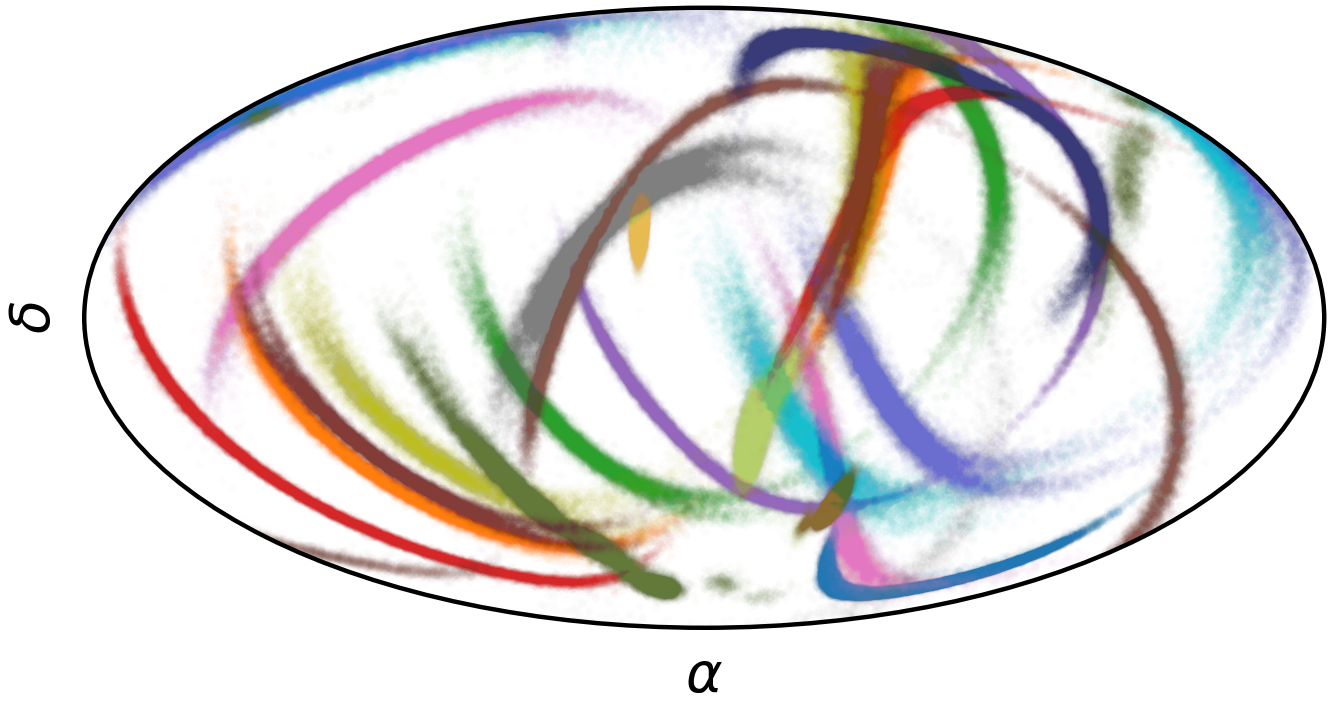}
    \includegraphics[width=.97\linewidth]{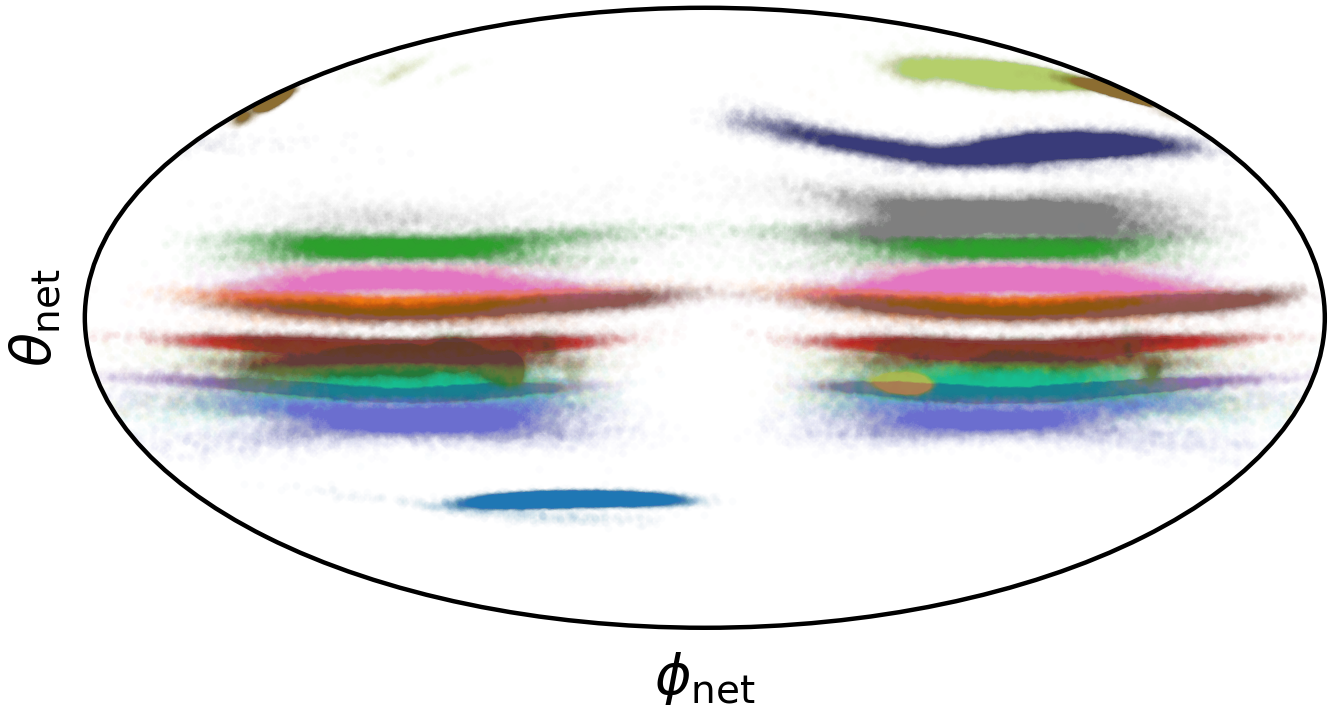}
    \caption{Sky location of events in the first two observing runs, expressed in terms of right ascension and declination (top) or network-based angles $\thetanet, \phinet$ defined in Fig.~\ref{fig:ligo_angles} (bottom).
        $\thetanet$ is typically well constrained and largely uncorrelated with $\phinet$.
        The signals are clearly clustered near $\los = \pm \bm{\hat y}_{\rm net}$, at the two high-sensitivity lobes of the network antenna pattern.
        Accordingly, when two or less detectors dominate the information content, the azimuthal angle $\phinet$ can be significantly informed by the antenna-pattern dependent prior.}
    \label{fig:thetaphinet}
\end{figure}

Figure~\ref{fig:thetaphinet} shows that, unlike $\alpha$ and $\delta$, $\thetanet$ is typically well measured and largely uncorrelated with $\phinet$.
In fact, from Eqs.~\eqref{eq:dt} and \eqref{eq:dt_detectors} its expected uncertainty is
\begin{equation} \label{eq:dcosthetanet}
\begin{split}
    \Delta \cos\thetanet
        &\sim \frac{1}{2 \pi \sigma_f \tau_{k_0 k_1}}
            \sqrt{\rho_{k_0}^{-2} + \rho_{k_1}^{-2}} \\
        &= 0.016 \cdot\frac{\SI{100}{\hertz}}{\sigma_f}
            \cdot\frac{\tau_{\rm HL}}{\tau_{k_0 k_1}}
            \cdot\frac{10}{\rho_{k_1}}
            \sqrt{1+\frac{\rho_{k_1}^2}{\rho_{k_0}^2}}
\end{split}
\end{equation}
for events observed in two detectors.
Each pair of detectors constrains the source location to a narrow ring in the celestial sphere.
This degeneracy is broken when the signal is prominent in three or more detectors.
The distribution of $\phinet$ is often multimodal, and for this reason we will ultimately use a modified azimuthal coordinate $\phinethat$ instead (see Sec.~\ref{ssec:symmetries}).

For single-detector events, a coordinate system aligned with the arms of the detector is more convenient.

\subsection{Arrival phase}
\label{ssec:phase}

\begin{figure}
    \centering
    \includegraphics[width=.6\linewidth]{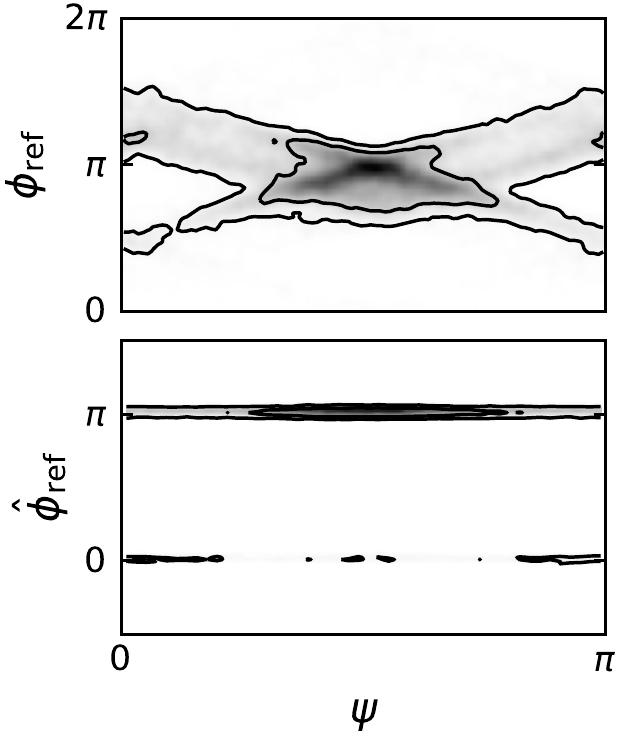}
    \caption{Unlike the reference orbital phase $\phiref$ (top), the coordinate $\phirefhat$ is well measured (bottom), since it uniquely determines the observable phase of arrival at the reference detector $\varphi_{k_0}$.
    There remains, however, a discrete degeneracy between solutions separated by $\pi$ that we treat in Sec.~\ref{sec:multimodality}.}
    \label{fig:phi}
\end{figure}

We now seek a parametrization of the form in Eq.~\eqref{eq:shear} that expresses the well-measured arrival phase $\varphi_{k_0}$ in terms of our new coordinates.
As is clear from Eq.~\eqref{eq:varphi_k}, such a reparametrization must involve $\iota$, $\los$, $\psi$, $\phiref$ and $\trefdet$.
To achieve this, we replace the coalescence phase $\phiref$ by another $2\pi$-periodic coordinate
\begin{align}
    &\phirefhat(\phiref, \iota, \los, \psi, \trefdet) \nonumber\\
    &\qquad\coloneqq \phiref + \frac{
        \arg{R_{k_0}(\iota, \los, \psi)}- 2\pi \fbar^{\rm ML}_{k_0} \trefdet - \varphi_{k_0}^{\rm ML}}{2} \label{eq:phirefhat_definition} \\ 
    &\qquad\equiv \frac{\varphi_{k_0} - \varphi_{k_0}^{\rm ML}}{2}, \label{eq:phirefhat_interpretation}
\end{align}
where we choose the constants $\fbar^{\rm ML}_{k_0}$ and $\varphi_{k_0}^{\rm ML}$ as the first frequency moment and arrival phase at the dominant detector for the (approximate) maximum likelihood source configuration.
$\phirefhat$ can be interpreted as the deviation of the coalescence phase $\phiref$ from the value that would make the arrival phase at the reference detector equal to $\varphi_{k_0}^{\rm ML}$.
Therefore, its posterior distribution should have a peak near 0.
The factor of $1/2$ in Eq.~\eqref{eq:phirefhat_interpretation} reflects the fact that the phase of the dominant quadrupolar radiation advances by twice the angle under an azimuthal rotation of the source.
We thus expect the $\phirefhat$ posterior to have a second mode near $\pi$.
By virtue of Eqs.~\eqref{eq:dphi} and \eqref{eq:phirefhat_interpretation}, these modes are largely uncorrelated with all other parameters and have widths $\sim 1/ (2 \rho_{k_0})$.
Radiation harmonics with odd values of $m$ break the symmetry between these two modes.
Figure~\ref{fig:phi} shows that $\phirefhat$ is indeed much better measured and less correlated than $\phiref$.
The transformation $\phiref\to\phirefhat$ has unit Jacobian.

Despite removing correlations, the definition of $\phirefhat$ in Eq.~\eqref{eq:phirefhat_definition} has two undesirable properties for sampling: its posterior is bimodal and, for waveforms with higher modes, discontinuous at the branch cut of $\arg R_{k_0}$.
We will solve both problems in Sec.~\ref{ssec:folding}.

\subsection{Reference frequency}
\label{ssec:fref}

Source parameters that are not conserved throughout the inspiral and merger need to be specified at a reference point in time, typically in terms of the instantaneous frequency $\fref$ of the quadrupole radiation.
Such parameters include orbital phase and, for precessing binaries, spin components, orbital inclination and polarization.
These are best constrained near frequencies where the signal is prominent; specifying them at a far-removed $\fref$ may introduce large correlations with other source parameters~\cite{Farr2014, Varma2021}.
This effect is most important for the orbital phase, which in turn couples to the in-plane spin azimuths if they are defined relative to the orbital separation vector (we revisit this in Sections \ref{ssec:inplane_spins} and \ref{ssec:azimuth}).
We will choose the reference frequency to be $\fref = \overline f_{k_0}^{\rm ML}$; from Eq.~\eqref{eq:fmoment}, this is the frequency weighted by the squared signal-to-noise ratio, and hence it will naturally be within the detector band.

\subsection{Aligned spin components}
\label{ssec:aligned_spins}

The effect of spins on the phase evolution of the inspiraling binary is largely dominated by the effective spin parameter
\begin{equation}
    \label{eq:chieff}
    \chieff = \frac{\sonez + q \, \stwoz}{1 + q},
\end{equation}
where $\sonez, \stwoz$ are the components of the constituent (dimensionless) spins parallel to the orbital angular momentum, and $q = m_2 / m_1 \leq 1$ is the mass ratio.
It is therefore convenient to use $\chieff$ directly as a spin coordinate.
The Jacobian of the transformation to Cartesian spin components is nontrivial \cite{Callister2021}, however, note that the prior on spins is much less certain than that on extrinsic parameters.
Our approach is to specify the prior in terms of $\chieff$, which means we do not need to compute the Jacobian.
We choose a uniform prior between $\pm 1$, which has the advantage that it does not vanish for any value of the observable $\chieff$.
In order to specify the two orbit-aligned spin components, we also sample over their (typically) poorly measured difference, whose conditional prior we choose to be uniform within the range allowed by the cosmic censorship conjecture, $\abs{\chi_{1, 2}} < 1$ (see appendix~\ref{app:reference}).
The prior induced on individual spin components is shown in \citet[][Fig.~1]{Huang2020}.

For systems with well measured $\chieff$, a common alternative description based on the spin magnitudes and tilts can suffer from correlations between these four variables.

\citet{Morisaki2020, Lee2022} proposed a different parametrization which also removes the correlation of the aligned spins with the masses. We intend to implement this in the future.

\subsection{In-plane spin components}
\label{ssec:inplane_spins}

In Sec.~\ref{ssec:aligned_spins} we have argued that $\bm L$ defines one preferred axis to parametrize the spins, as their aligned components affect the evolution of the orbital phase.
Describing the spin azimuths requires us to define a second axis, for which two choices are common in the literature: the orbital separation vector at $\fref$, or the direction of propagation $\propagation \equiv - \los$.
In line with \citet{Farr2014}, here we will advocate the latter choice.

Misaligned spins cause the orbital angular momentum $\bm L$, as well as the spins $\bm S_1, \bm S_2$, to precess about the total angular momentum $\bm J$, whose direction is stable \cite{Apostolatos1994, Kidder1995}.
This causes the inclination of the orbit seen from Earth, determined by $\bm {\hat L} \cdot \propagation$, to continuously change.
As a consequence, in addition to the usual cycles at twice the orbital frequency, misaligned-spin waveforms exhibit amplitude modulations at the slower precession rate.
This separation of timescales means that the  precession dynamics can be described using orbit-averaged equations, thereby decoupling it from the orbital phase.

\begin{figure}
    \centering
    \includegraphics[width=.6\linewidth]{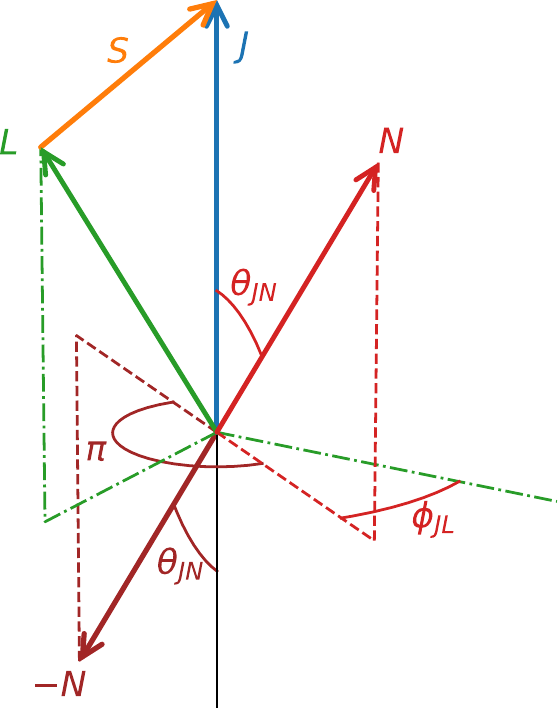}
    \caption{Angles describing relative orientation between angular momenta and the direction of wave propagation \cite{Farr2014}.
    In terms of the observable precession cycles, changing $\bm {\hat N} \mapsto -\bm {\hat N}$ (i.e.\ inverting $\cos\thetajn$ and adding $\pi$ to $\phi_{JL}$) is a symmetry that can cause a discrete degeneracy.
    We return to this point in Sec.~\ref{ssec:symmetries}.}
    \label{fig:spins}
\end{figure}

Thus, the size and peak frequencies of the amplitude modulations are governed by the orbital and spin angular momenta at $\fref$ relative to $\bm {\hat N}$, as these determine the evolution of the inclination.
On the other hand, per Eq.~\eqref{eq:varphi_k} the phase $\varphi_k$ of the ``carrier'' wave is controlled by an approximately degenerate combination of $\phiref, \psi, \iota, \los, t_c$.
The practical consequence of this degeneracy is that $\phiref$, which defines the azimuth about $\bm L$ between the orbital separation and the direction of propagation, is poorly measured.
In other words, a change in $\phiref$ can be compensated by adjusting e.g.\ $\psi, \dl$ to keep $\phirefhat, \dhat$ fixed, but only if the spins are held constant relative to $\propagation$.
If, instead, the spins are rigidly rotated with the binary, the peak frequencies of the observed amplitude modulations would shift in a way that cannot be compensated by adjusting other parameters.
Thus, using the orbital separation to define the origin of spin azimuths would introduce in the parametrization a spurious coupling between the observable precession and orbital cycles.
Following \citet{Farr2014}, we will use the angles $\thetajn, \phi_{JL}, \phi_{12}$ to describe the inclination of the orbit and the spin azimuths.
$\thetajn$ and $\phi_{JL}$ are illustrated in Fig.~\ref{fig:spins}: they define the direction of wave propagation $\propagation$ given the total and orbital angular momenta $\bm J, \bm L$ and irrespective of the orbital separation vector.
$\phi_{12}$ is defined as the difference between the primary and secondary spin azimuths about $\bm L$.
$\phi_{JL}$ posteriors are often bimodal, and for this reason we will eventually replace $\phi_{JL}$ by a modified spin azimuth $\phijlhat$ in Sec.~\ref{ssec:symmetries}.

Our parametrization of the spins is more akin to cylindrical than spherical coordinates, in the sense that we favor the orbit-aligned spin components over the zenithal angles.
The description of spins is completed by specifying the in-plane spin magnitudes $\soner, \stwor$.

\section{Mitigating multimodality}
\label{sec:multimodality}

Oftentimes, the posterior distribution of gravitational wave source parameters is multimodal.
Such distributions are more challenging for stochastic samplers, which might occasionally misestimate the relative weights of the modes or miss some of them altogether.

In Sec.~\ref{ssec:symmetries} we identify four approximate discrete symmetries that are responsible for frequent multimodality in the orbital phase, polarization, inclination and sky location parameters.
In Sec.~\ref{ssec:folding} we introduce ``folding'': a simple algorithm to exploit the knowledge of the underlying approximate symmetries in order to robustly sample this type of multimodal probability distribution.
Depending on the source properties, the detector network and the signal-to-noise ratio, this procedure can generically reduce the number of disjoint modes by a factor of up to 8.

\subsection{Approximate discrete symmetries}
\label{ssec:symmetries}

In Sec.~\ref{ssec:phase} we already encountered an approximate discrete symmetry
\begin{equation}
    \phirefhat \mapsto \phirefhat + \pi, \label{eq:phi_symmetry}
\end{equation}
(Fig.~\ref{fig:phi}) which is exact for waveforms with only even values of $m$, in particular the dominant mode $(\ell, \abs{m}) = (2, 2)$.

A similar symmetry exists for the polarization
\begin{equation}
    \psi \mapsto \psi + \frac\pi2 \label{eq:psi_symmetry}
\end{equation}
at constant $\phirefhat$, which (per Eq.~\eqref{eq:phirefhat_definition}) entails a simultaneous change by $\pi/2$ in $\phiref$.
This symmetry arises because, under the transformation \eqref{eq:psi_symmetry}, the antenna coefficients $F^+_k, F^\times_k$ change sign \cite{Whelan2013}; meanwhile, the $\pi/2$ rotation of $\phiref$ inverts the sign of the waveforms $h^+_{\ell m}, h^\times_{\ell m}$ for modes with $\abs{m}=2$.
These two sign flips cancel, leaving the measurable responses $h_k = F^+_k h^+ + F_k^\times h^\times$ invariant.
The symmetry becomes exact for waveforms with $\abs{m}=2$.
Usually, this discrete degeneracy does not produce disjoint modes in the posterior, because the uncertainty in $\psi$ is large enough that the two solutions remain connected (as an extreme example, $\psi$ is a dummy parameter for face-on, aligned-spin waveforms).
Still, $\psi$ posteriors generally do exhibit this symmetry.

We identify two other approximate discrete symmetries.
In Sec.~\ref{ssec:amplitude}--\ref{ssec:phase} we derived coordinates $\dhat,\allowbreak \phirefhat,\allowbreak \trefdet,\allowbreak \thetanet$, which determine the amplitude, phase and time of arrival at the leading detector, and arrival time at the second detector.
These variables typically capture most of the information on extrinsic parameters from the likelihood, and consequently tend to be tightly constrained.
Conversely, the remaining extrinsic parameters $\thetajn, \phinet, \psi$ have a smaller effect on the likelihood.
As a result, they can exhibit large degeneracies---some of them discrete, leading to multiple modes.
Indeed, due to the geometry of the Hanford--Livingston network, for these detectors both the prior and likelihood are approximately symmetric under either of the following discrete transformations (at fixed $\dhat, \phirefhat, \trefdet, \thetanet$):
\begin{subequations}
\label{eq:symmetries}
\begin{align}
    \phinet &\mapsto -\phinet, \label{eq:over_under} \\
    (\phinet, \cos\thetajn, \phi_{JL}) &\mapsto (\phinet+\pi, -\cos\thetajn, \phi_{JL}+\pi).
    \label{eq:faceon_faceoff}
\end{align}
\end{subequations}

\begin{figure}
    \centering
    \includegraphics[width=.9\linewidth]{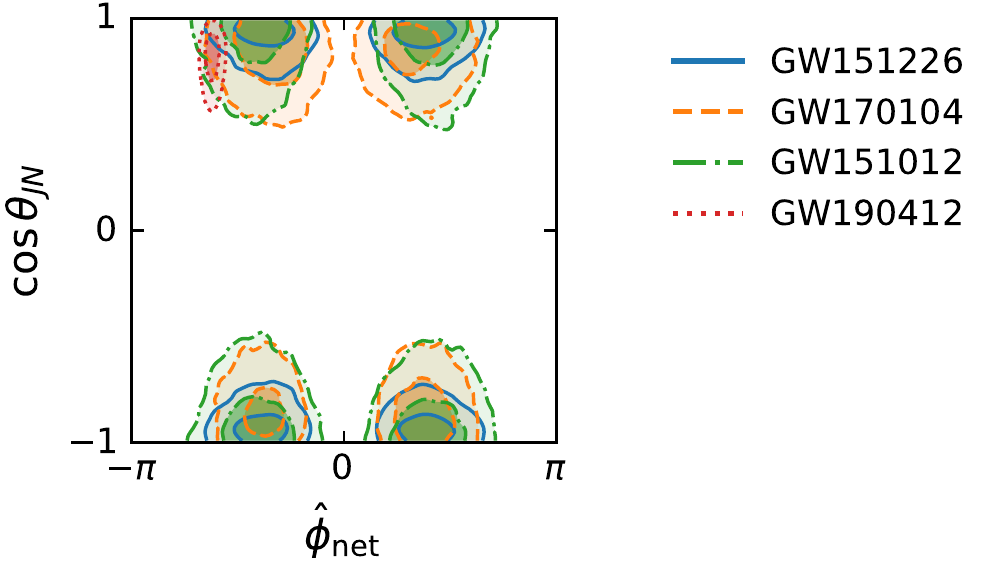}
    \caption{Approximate discrete symmetries are frequently responsible for multiple modes in the distribution, especially for the Hanford--Livingston network due to its peculiar geometric configuration.
    The plot shows examples of posteriors for the cosine of the inclination and a coordinate $\phinethat$ (Eq.~\eqref{eq:phinet_hat}) that describes the line-of-sight azimuth along a ring of constant time delay.
    Additional detectors and higher modes can break these symmetries, as for GW190412 \cite{Abbott2020_GW190412}.}
    \label{fig:symmetries}
\end{figure}

The manifestation of these symmetries in the posterior is shown in Fig.~\ref{fig:symmetries}.
We can understand them intuitively as follows.
By design, the Hanford and Livingston detectors are nearly coaligned (plus a $\pi/2$ rotation in the horizontal plane, which simply adds a phase difference of $\pi$; see Fig.~\ref{fig:ligo_angles}).
For perfectly coaligned detectors, the amplitude and phase at the second detector are determined by those at the first; in this sense their measurement does not provide new constraints, and the likelihood is largely independent of $\phinet, \iota, \psi$.
These parameters are therefore significantly informed by the prior.
The prior has nontrivial structure in these variables because the distance, and thereby the observable volume, depends on $\iota, \phinet, \psi$ at constant $\dhat, \thetanet$ (which are fixed by the likelihood).
For example, the uniform prior in luminosity volume $\pi(\dl)\propto\dl^2$ is (by Eqs.~\eqref{eq:dhat} and \eqref{eq:dhat_jacobian})
\begin{equation}
    \label{eq:dhat_prior}
    \pi(\dhat \mid \mchirp, \iota, \thetanet, \phinet, \psi)
    \propto \dhat^2 \mchirp^{5/2}
        \abs{R_{k_0}(\iota, \thetanet, \phinet, \psi)}^3,
\end{equation}
proportional to the cube of the absolute value of the reference detector response.
This term has four peaks as a function of $\iota, \phinet$: near $\cos\iota = \pm 1$ (Eq.~\eqref{eq:response}) and $\phinet = \pm\pi/2$ (Fig.~\ref{fig:thetaphinet}); see also Fig.~\ref{fig:azimuth_symmetry}.
In other words, corresponding to a source face-on or face-off, and above or beneath the detector (constrained to the time-delay ring).
As a result, Hanford--Livingston posteriors typically exhibit four modes at these configurations.
We can see that both transformations \eqref{eq:symmetries} map an overhead source to one underfoot, i.e., they send $\los \cdot \bm{\hat y}_{\rm net} \equiv n_y \mapsto -n_y$.
The second transformation, Eq.~\eqref{eq:faceon_faceoff}, also flips the inclination.
All four configurations can be reached by applying combinations of these two transformations.

The transformation $\cos\thetajn, \phi_{JL}\mapsto -\cos\thetajn, \phi_{JL}+\pi$ in Eq.~\eqref{eq:faceon_faceoff} corresponds to inverting the direction of propagation $\propagation \mapsto -\propagation$ in the frame of the binary.
In other words, it can be interpreted as observing the system from the antipodal direction.
To the extent that the source can be modeled as a precessing quadrupole, this simply reverses the handedness of the gravitational wave and cannot be distinguished with coaligned detectors.
In reality, due to the curvature of Earth the detectors are not perfectly coaligned.
As a result, the phase of arrival at the second detector is not completely determined by that at the first detector and its measurement has some constraining power.
Whether it is advanced or retarded depends on the handedness of the elliptically polarized wave and the relative orientation of the two detectors as viewed from the source (the ``projected detector tensors'' \cite{Whelan2012}).
For example, with the system defined in Fig.~\ref{fig:ligo_angles}, the projected Hanford detector appears rotated counter-clockwise relative to Livingston when seen from the $x_{\rm net} > 0$ hemisphere, but clockwise from $x_{\rm net} < 0$.
Thus, the phase at Hanford would be advanced relative to Livingston for a right-polarized wave from $n_x < 0$, or left-polarized from $n_x > 0$ (and vice versa), making these solutions degenerate.
Thus, $n_x$ has to be inverted simultaneously for the transformation $\propagation \mapsto - \propagation$ to be a good symmetry of the likelihood.
This is implemented by sending $\phinet \mapsto \phinet + \pi$ in Eq.~\eqref{eq:faceon_faceoff}.

Incidentally, since the arrival phase difference also involves the time delay, this effect induces an observable correlation between $\thetanet$ and $\phinet$.
We provide a more quantitative treatment of all this in Appendix~\ref{app:symmetries}.

In order to simplify the transformation \eqref{eq:faceon_faceoff}, we will define two $2 \pi$-periodic coordinates
\begin{align}
    \label{eq:phinet_hat}
    \phinethat &\coloneqq \begin{cases}
        \phinet & \text{if $\cos\thetajn < 0$} \\
        \phinet + \pi & \text{else}
        \end{cases} \\
    \label{eq:phijl_hat}
    \phijlhat &\coloneqq \begin{cases}
        \phi_{JL} & \text{if $\cos\thetajn < 0$} \\
        \phi_{JL} + \pi & \text{else}
        \end{cases}
\end{align}
to replace $\phinet$ and $\phi_{JL}$ in the characterization of the sky location and spin azimuths, respectively.
Both transformations have unit Jacobian.
With these coordinates, the approximate symmetries in Eq.~\eqref{eq:symmetries} become one-parameter reflections:
\begin{subequations}
\label{eq:symmetries_v2}
\begin{align}
    \phinethat &\mapsto -\phinethat, \label{eq:over_under_v2} \\
    \cos\thetajn &\mapsto -\cos\thetajn.
    \label{eq:faceon_faceoff_v2}
\end{align}
\end{subequations}
$\phijlhat$ has a simple interpretation as the azimuth about $\bm J$ between $\bm L$ and the unsigned direction of propagation, so that $\propagation \mapsto -\propagation$ leaves $\phijlhat$ invariant.
In Sec.~\ref{ssec:azimuth} we will show that this azimuth can be remarkably well measured.

The Virgo detector does not have such special alignment.
If a signal is loud in Virgo and at least one LIGO detector, the symmetries in Eq.~\eqref{eq:symmetries_v2} are typically broken and the posterior distribution in these variables is unimodal.

\subsection{Folding algorithm}
\label{ssec:folding}

\begin{figure}
    \centering
    \includegraphics[width=\linewidth]{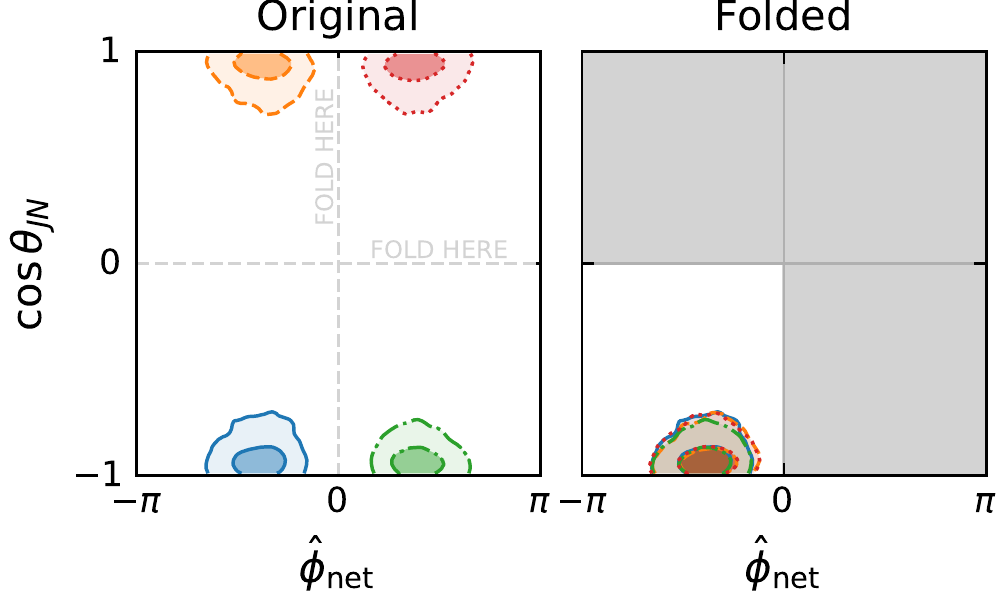}
    \caption{Folding algorithm.\textit{ Left:} the posterior has multiple modes (shown in different colors) due to approximate symmetries that are known in advance.\textit{ Right:} we can ``fold'' the distribution (sum its appropriately transformed modes) to make it unimodal.
    We sample the folded distribution and reconstruct the original in postprocessing.}
    \label{fig:folding}
\end{figure}

Having identified the approximate discrete symmetries responsible for multimodality, we now introduce ``folding'', an algorithm to exploit this structure when sampling.
The basic idea is illustrated in Fig.~\ref{fig:folding}, beginning with sampling from the probability distribution marginalized over the discrete approximate-symmetry transformations, and then sampling over the set of transformations in postprocessing to undo the marginalization.

The four approximate symmetry transformations in Eqs.~\eqref{eq:phi_symmetry}, \eqref{eq:psi_symmetry}, \eqref{eq:over_under_v2}, \eqref{eq:faceon_faceoff_v2} allow us to divide the phase space into $2^4=16$ sectors in which the posterior has similar behavior.
This is illustrated in the left panel of Fig.~\ref{fig:folding} for the four quadrants of $\cos\thetajn, \phinethat$ space, which have similar solutions by virtue of the approximate symmetries in Eq.~\eqref{eq:symmetries_v2} (the other two folded dimensions $\phirefhat, \psi$ are not shown).
We pick one of these sectors, as highlighted in the right panel, which we call the ``folded space''.
The full space can be recovered from the folded space via $2^4$ discrete mappings $\{\sigma_1, \ldots, \sigma_{16}\}$ that either do or do not apply each of the transformations \eqref{eq:phi_symmetry}, \eqref{eq:psi_symmetry}, \eqref{eq:over_under_v2}, \eqref{eq:faceon_faceoff_v2}, respectively.
If the symmetries were perfect, it would suffice to draw samples on the folded space, then distribute them evenly onto all sectors using these mappings.
The symmetries in Section \ref{ssec:symmetries} are only approximate, since the detectors are not perfectly aligned, and general waveform models have additional effects like orbital precession and higher harmonics that are not modeled by Eq.~\eqref{eq:h22np}.
Hence we generalize this idea to relax the requirement of perfect symmetry, as follows:
\begin{enumerate}
    \item Define the folded distribution
    \begin{equation}\label{eq:folding}
        \tilde P(\tilde\theta) =
            \sum_{i=1}^{2^N} P(\sigma_i(\tilde\theta)),
    \end{equation}
    where $\tilde\theta$ is a set of parameters in the folded space, $N=4$ is the number of folded parameters, $\{\sigma_i\}$ are the $2^N$ discrete mappings and $P$ is the posterior distribution in the full space.
    The number of modes of $\tilde P$ can be smaller than that of $P$, as shown in the right panel of Fig.~\ref{fig:folding}, by a factor of up to $2^N$.

    \item Draw ``folded'' samples from this simpler distribution, $\{\tilde \theta^j\} \sim \tilde P$, using a traditional sampler.

    \item\label{step:resampling} Assign each folded sample to a sector: for each $\tilde\theta^j$, draw $\theta^j$ from $\{\sigma_1(\tilde\theta^j), \sigma_2(\tilde\theta^j), \ldots\}$ with relative probabilities $\{P(\sigma_1(\tilde \theta^j)), P(\sigma_{2}(\tilde \theta^j)), \ldots\}$.
    The set $\{\theta^j\}$ is distributed according to $P(\theta)$.
\end{enumerate}

At first glance, Eq.~\eqref{eq:folding} suggests that each evaluation of $\tilde P$ requires $2^N$ evaluations of $P$ and would therefore increase the computational cost by that factor.
However, since the folded parameters are extrinsic, the expensive computation of the waveform can be reused.
In the case at hand, changing the polarization and sky location requires recomputing antenna factors $F_+, F_\times$ and time delays, and changing the phase requires recomputing spherical harmonic phases $e^{i m \phiref}$.
Even in a general case, all $2^N$ evaluation points are known simultaneously, facilitating vectorization and parallelization.
These considerations make the cost of computing $\tilde P$ similar to that of $P$.
Likewise, generating $\theta^j$ from $\tilde \theta^j$ in step \ref{step:resampling} above does not require additional computations of $P$, because these values can be stored along with $\tilde \theta^j$ during the sampling process, and the assignment itself is computationally very cheap.
In practice, the number of evaluations required for convergence is much smaller for distributions with less modes, making this method advantageous for both robustness and efficiency.

We emphasize that the folding procedure is not an approximation: it still gives the correct answer if the distribution $P$ is not symmetric at all (in that case, it just does not provide any advantage).

In Sec.~\ref{ssec:phase} we had mentioned that the $\phirefhat$ posterior is bimodal and discontinuous.
In contrast, per Eqs.~\eqref{eq:phirefhat_definition}, \eqref{eq:phi_symmetry} and \eqref{eq:folding} the folded posterior is unimodal and continuous, because both sides of the branch cut are summed together.
In a similar way, the discontinuities introduced in Eqs.~\eqref{eq:phinet_hat} and \eqref{eq:phijl_hat} are absent in the folded posterior since they happen at the fold $\cos\thetajn = 0$.

\begin{figure}
    \centering
    \includegraphics[width=\linewidth]{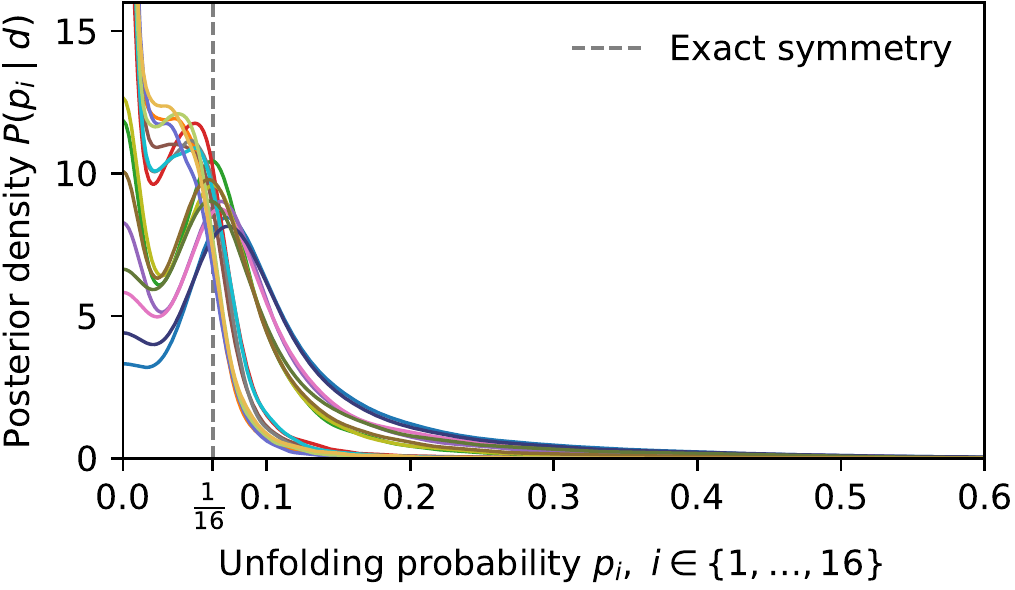}
    \caption{Distribution of probabilities $p_i$ with which folded samples are assigned to the 16 sectors during step \ref{step:resampling}, for GW151012.
    All peak near 1/16, demonstrating that the transformations \eqref{eq:phi_symmetry}, \eqref{eq:psi_symmetry}, \eqref{eq:over_under_v2}, \eqref{eq:faceon_faceoff_v2} are approximate symmetries of the full 15-dimensional space.}
    \label{fig:folding_probabilities}
\end{figure}

Finally, one might worry that Fig.~\ref{fig:folding} merely shows that the marginalized 2-dimensional posterior is approximately symmetric, while the folding algorithm is efficient only if the full 15-dimensional posterior is symmetric.
It could be possible that, although this projection looks symmetric, the mappings we proposed were incorrect descriptions of the symmetry in the full space.
In Fig.~\ref{fig:folding_probabilities} we show histograms of the unfolding probabilities $p_i \propto P(\sigma_i(\tilde\theta))$ used in step \ref{step:resampling}, which have unit sum by construction.
In the limit that the transformations $\sigma_i$ were perfect symmetries of $P$, these probabilities would be 1/16.
Conversely, if they were not good symmetries, these probabilities would be very nearly 0 or 1.
We find that they are near $1/16$, confirming that the transformations we identified are indeed approximate symmetries of the full space.
Whether all or some of the modes are present for any particular event depends on the source parameters, the network configuration and the signal-to-noise ratio.

An alternative to the folding algorithm, which is similar in spirit, is to specify custom jump proposals that follow the approximate symmetry transformations and effectively ``connect'' the modes in the sampling process.
This approach has been used to mitigate both degeneracy and multimodality \cite{Veitch2015, Cornish2015, Ashton2021}, albeit for a different set of symmetries (that includes the transformation in Eq.~\eqref{eq:phi_symmetry}).
The choice of sampling proposal can significantly impact the efficiency of Monte Carlo methods, see e.g.~\cite{Sharma2017,Hogg2018} and references therein.

\section{Results}
\label{sec:results}

We implement the coordinate transformations described in Sec.~\ref{sec:coordinates} and the folding algorithm introduced in Sec.~\ref{sec:multimodality} in a Python package, which we make publicly available at \url{https://github.com/jroulet/cogwheel}.
Apart from those improvements, the \texttt{cogwheel} code utilizes the relative-binning algorithm \cite{Cornish2013, Zackay2018, Cornish2021} with higher-order modes \cite{Leslie2021} to accelerate likelihood evaluations.
It interfaces with third-party routines for downloading public data (GWOSC~\cite{Abbott2021_GWOSC}, \texttt{GWpy}~\cite{Macleod2021}), generating waveforms (\texttt{lalsuite}~\cite{lalsuite}) and sampling distributions (\texttt{PyMultiNest}~\cite{Buchner2014, Feroz2009}, \texttt{dynesty}~\cite{Speagle2020}).

In this section, we summarize the improvements brought by the coordinate transformations and folding technique in terms of efficiency and robustness of the inference.
Throughout, we model waveforms using \texttt{IMRPhenomXPHM} with next-to-next-to-leading-order (NNLO) post-Newtonian precession \cite{Pratten2021}, and we sample distributions with \texttt{PyMultiNest}~\cite{Buchner2014,Feroz2009}.

\begin{figure*}
    \centering
    \includegraphics[width=.43\linewidth]{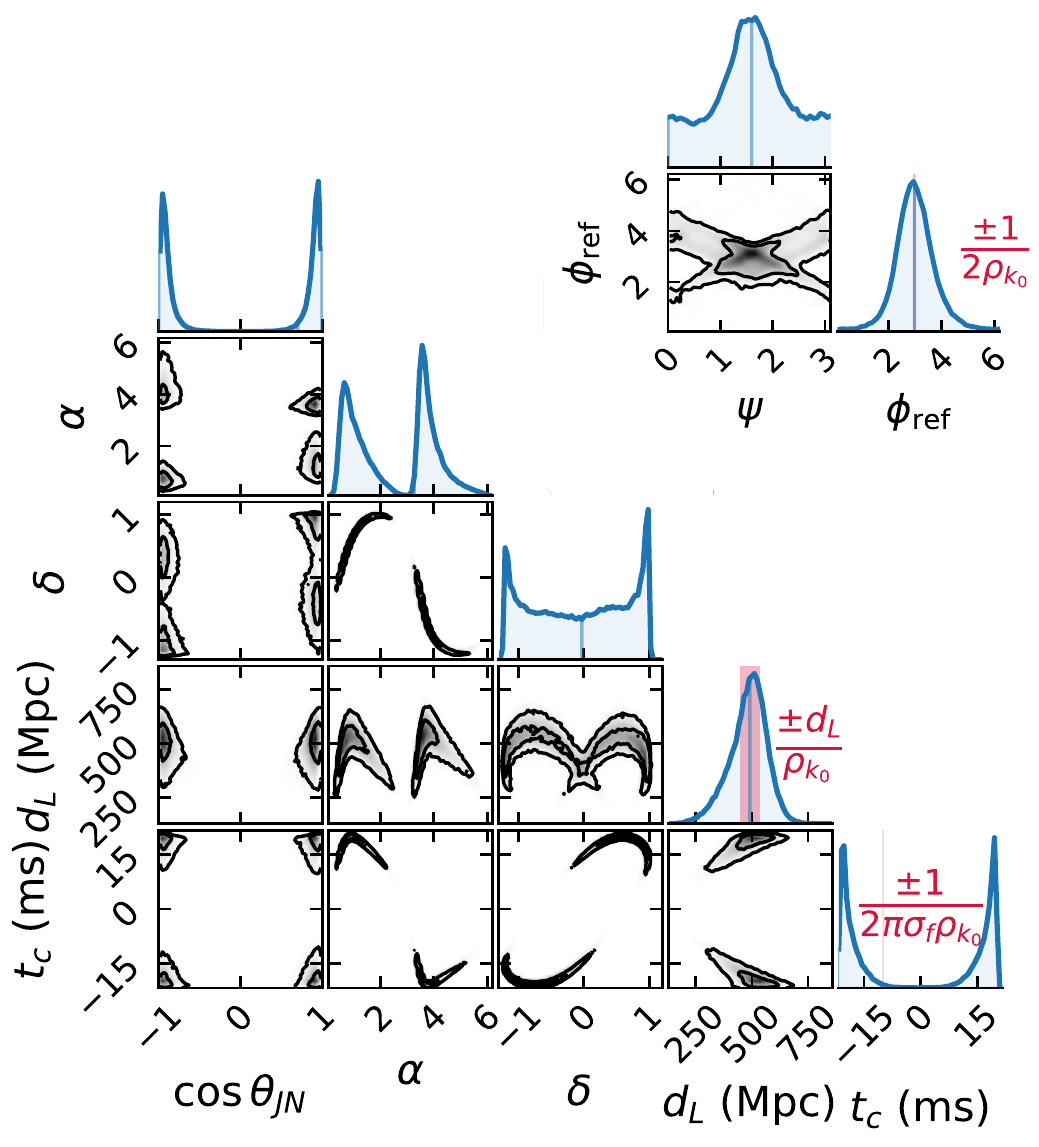}\hfill
    \includegraphics[width=.56\linewidth]{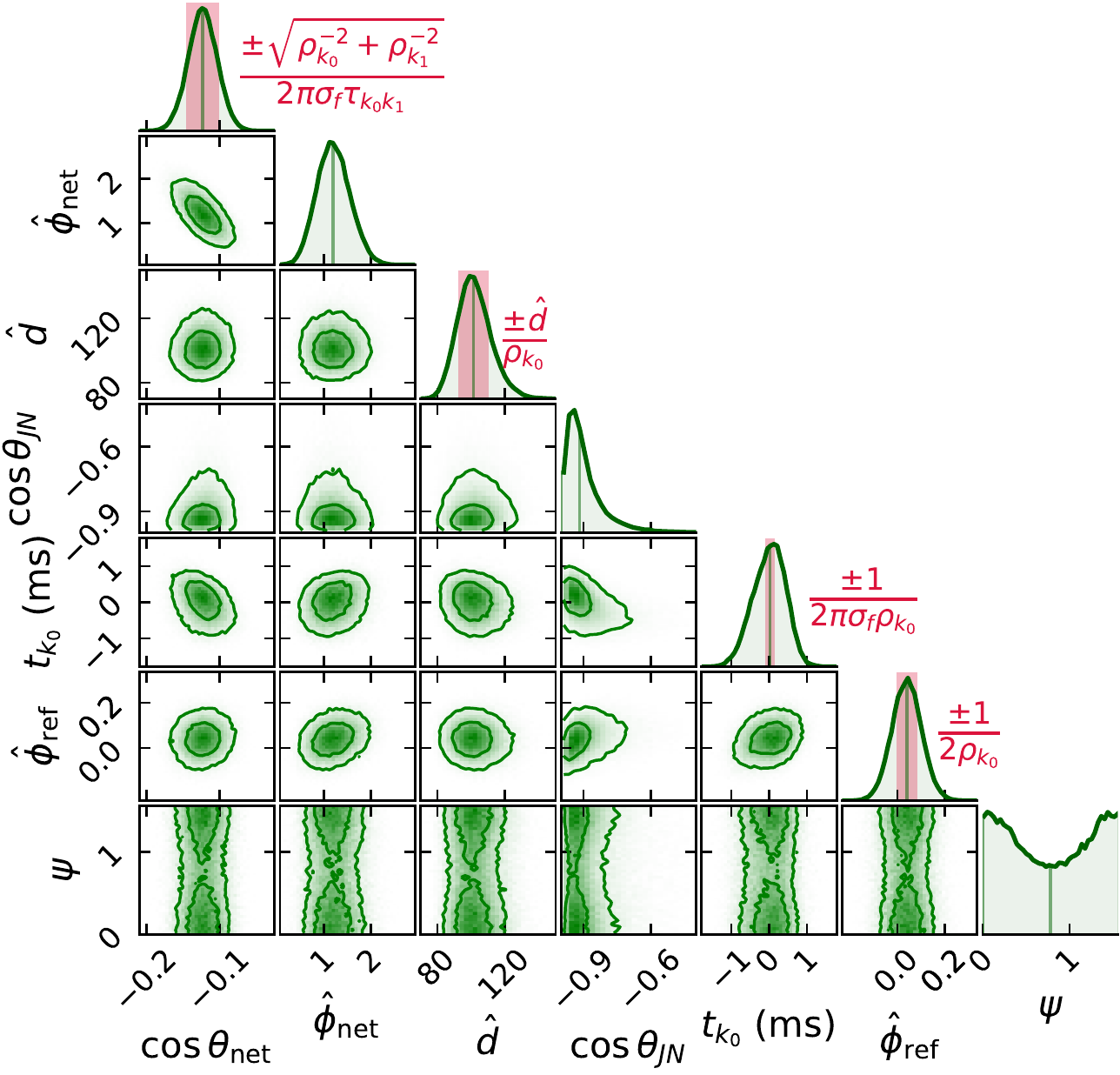}
    \caption{Marginalized posterior for the extrinsic parameters of GW151226 in terms of standard coordinates (left) or the folded coordinates advocated in this work (right), see Sections~\ref{sec:coordinates} and \ref{sec:multimodality}.
    The complex structures on the left are largely absent on the right, making the latter system better suited for sampling.
    Crimson bars show the standard deviation expected given the signal-to-noise ratio (Eqs.~\eqref{eq:da}--\eqref{eq:dt}, \eqref{eq:dcosthetanet}), discrepancies with the realized uncertainties are due to correlations with the marginalized parameters.
             }
    \label{fig:before_after}
\end{figure*}

Figure~\ref{fig:before_after} summarizes how our choice of coordinates, combined with the folding method, naturally describes the distribution of extrinsic parameters.
The uncertainties achieved in the one-dimensional marginal posteriors for the parameters controlling distance, arrival phase and arrival times are in much closer agreement with the expectations from Eqs.~\eqref{eq:da}--\eqref{eq:dt} and \eqref{eq:dcosthetanet}, shown by crimson bars.
That being said, Eq.~\eqref{eq:dt} significantly underestimates the uncertainty of the arrival time $\trefdet$.
This can be traced back to correlations with intrinsic parameters, most significantly chirp mass and in-plane spin magnitude (not shown).
The definition of arrival time for waveforms with different intrinsic parameters is somewhat arbitrary.
The convention adopted by the LIGO Algorithm Library is to define the reference time when the amplitude of the strain is maximal \cite{Schmidt2017}.
In the \texttt{IMRPhenomX} family of waveform models this is implemented through parametric fits \cite{GarciaQuiros2020} that are accurate to $\sim\SI{1}{\milli\second}$ \cite{Pratten_pc}, which is consistent with the precision of the arrival time measurement achieved in Fig.~\ref{fig:before_after}.

\subsection{Performance of stochastic sampling}
\label{ssec:performance}

We compare the performance of parameter estimation runs using the coordinate system presented in Sec.~\ref{sec:coordinates} (coupled to the folding algorithm of Sec.~\ref{ssec:folding}) against an ``unoptimized'' system that uses $
(\cos\thetanet,\allowbreak
\phinet,\allowbreak
\trefdet,\allowbreak
\cos\thetajn,\allowbreak
\phi_{JL},\allowbreak
\phi_{12},\allowbreak
\cos\theta_1,\allowbreak
\cos\theta_2,\allowbreak
\chi_1,\allowbreak
\chi_2,\allowbreak
\dl,\allowbreak
\phiref,\allowbreak
\psi,\allowbreak
\mchirp,\allowbreak
\ln q)$ as sampling coordinates and no folding.
This comparison assesses the impact of the methods new to this work within the \texttt{cogwheel} software, note that other state-of-the-art codes for parameter inference \cite{Veitch2015, Cornish2015, Lange2018, Ashton2019, Biwer2019, Ashton2021} use various other parametrizations and optimizations (marginalization, custom jump proposals, parallelization, etc.).
We adopt an isotropic, uniform-in-spin-magnitudes prior in both cases, and so for ease of implementation in the optimized case we use the cumulative of the prior on aligned spin components as sampling coordinate instead of the effective spin (cf.\ Sec.~\ref{ssec:aligned_spins}).
In both examples we use \texttt{PyMultiNest} and vary the number of live points in factors of two between 512 and \num{16384} (this is an internal feature of nested samplers; a higher number of live points typically achieves a better coverage of parameter space and produces more independent samples at a proportionally higher computational cost).
In all cases, we use a single computing core and identical environments to run each configuration, thus, differences in runtime are due to the algorithm.
We use GW151226 as a test case.

\begin{figure}
    \centering
    \includegraphics[width=\linewidth]{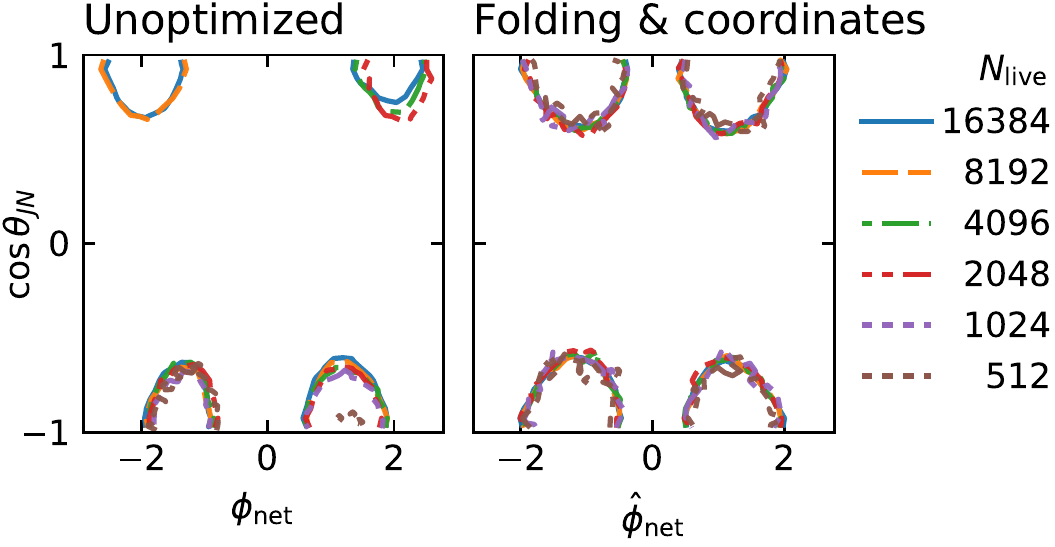}\\[4pt]
    \includegraphics[width=\linewidth]{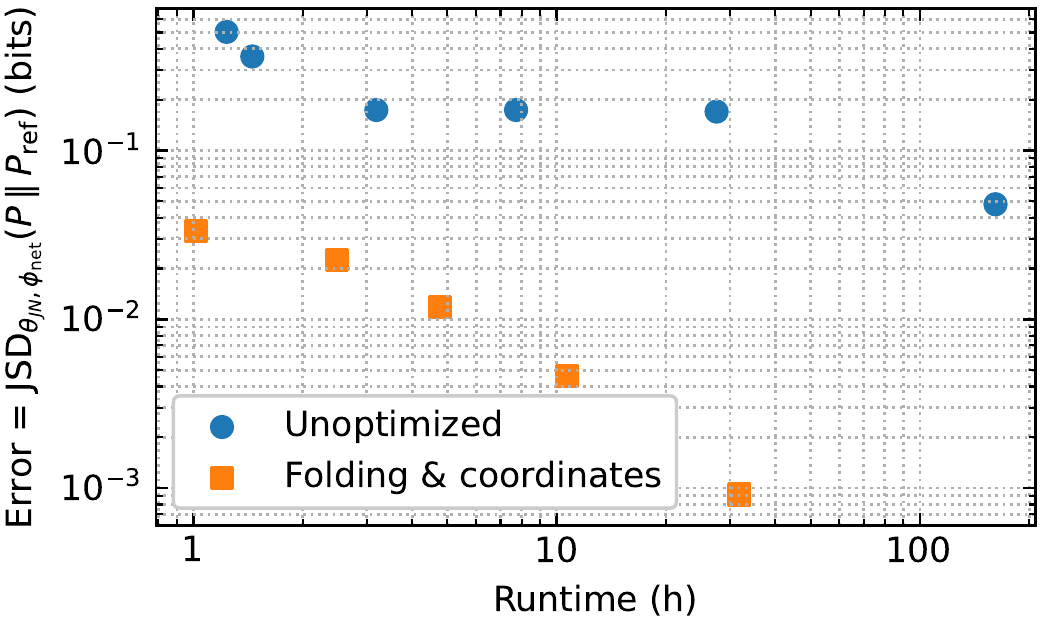}
    \caption{Improvement brought by our methods in terms of increased robustness to multimodality.
    We compare runs that do not use the methods new to this work (``Unoptimized'') to runs that do (``Folding \& coordinates'').
    Within each setting, runs differ in number of live points.
    \textit{Top:} Without folding, the sampler fails to find all modes of the posterior even when using a large number of live points.%
    \textit{ Bottom:} Error achieved in the posterior distribution versus runtime, as the number of live points is iteratively doubled.
    We use the Jensen--Shannon divergence with respect to the highest-resolution ``Folding \& coordinates'' run as our error measure, see Eq.~\eqref{eq:jsd} (the reference run is excluded from the figure).
    The unoptimized runs have a higher error due to missing or undersampled modes.}
    \label{fig:performance}
\end{figure}

We find that, despite nested sampling being generally well suited for multimodal problems, and despite using numerous live points, the sampler fails to find some modes of the posterior when folding is not used.
This is illustrated in the top panel of Fig.~\ref{fig:performance}: of the unoptimized runs, only the one with \num{16384} live points found all modes, and even in this case one of the modes was undersampled.
Conversely, all runs that used our coordinate system and folding successfully found all the modes.
To quantify the error this induces, we compare each posterior distribution $P$ to a reference answer $P_{\rm ref}$, chosen as that with the largest number of live points and using our coordinate system and folding algorithm.
Our measure of the error is the Jensen--Shannon divergence between each marginal $\thetajn, \phinet$ posterior and the reference distribution:
\begin{multline}
    \label{eq:jsd}
    {\rm JSD}_{\thetajn, \phinet}(P \parallel P_{\rm ref}) \\
    \coloneqq \frac 12 \iint \rmd \thetajn \, \rmd \phinet \left(
        P \log_2\frac PM + P_{\rm ref}\log_2 \frac{P_{\rm ref}}{M}
        \right)
\end{multline}
with $M = (P + P_{\rm ref}) / 2$.
We singled out $\thetajn, \phinet$ because this projection of the unoptimized posteriors is most clearly missing modes.
To evaluate Eq.~\eqref{eq:jsd}, we obtain $P(\thetajn, \phinet), P_{\rm ref}(\thetajn, \phinet)$ from the posterior samples with a kernel density estimation and perform a double quadrature, using \texttt{scipy} implementations \cite{Virtanen2020}.
We show this in the bottom panel of Fig.~\ref{fig:performance}: we see more than an order of magnitude improvement in terms of this error measure when we use our coordinate system and folding.

As expected, the runtime is approximately proportional to the number of live points, however, the unoptimized runs show larger fluctuations.
We attribute these to the stochastic nature of the sampling process, likely amplified by the fact the distribution is multimodal: the runtime can be significantly affected by whether, and when, a new mode is found by the sampler, especially when using few live points, for which the relative fluctuations are found to be largest.

\subsection{Measurability of spin azimuth}
\label{ssec:azimuth}

\begin{figure}
    \centering
    \includegraphics[width=\linewidth]{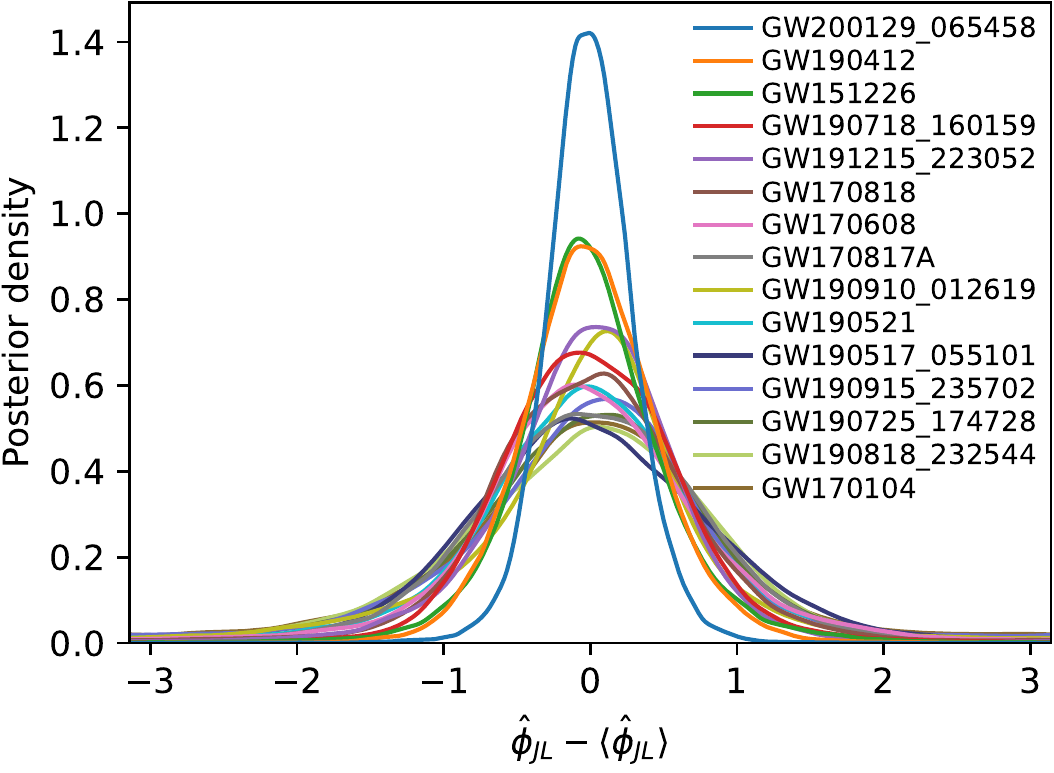}
    \caption{Posterior distributions of modified spin azimuth $\phijlhat$ (Eq.~\eqref{eq:phijl_hat}) for the 15 events with best constraints, under a prior isotropic in spins.
    For aligned-spin configurations, $\phijlhat$ becomes a dummy parameter, and therefore events with well-measured $\phijlhat$ have spins misaligned with the orbit.
    Posteriors have been artificially centered by subtracting their circular means, which have no astrophysical importance.}
    \label{fig:spin_azimuth}
\end{figure}

From the astrophysics standpoint, the most interesting result of this work is that the modified spin azimuth $\phijlhat$ can be measured remarkably well.
Figure~\ref{fig:spin_azimuth} shows its posterior distribution for the 15 events in the GWTC-3 and Institute for Advanced Study's catalogs with tightest bounds.
We see that it is not uncommon for this parameter to be significantly constrained.
Recall that $\phijlhat$ describes the azimuth about $\bm J$ between the orbital angular momentum $\bm L$ and the location of the detector (Eq.~\eqref{eq:phijl_hat}, Fig.~\ref{fig:spins}).
Due to isotropy, all possible directions $\propagation$ towards the detector are equivalent a priori, and thus the prior on $\phijlhat$ is uniform.
Moreover, for aligned-spin configurations we have $\bm L \parallel \bm J$, so $\phijlhat$ has no effect on the waveform and therefore the likelihood is independent of $\phijlhat$.
It follows that any departure from a flat posterior originates from the presence of misaligned spins in the waveform model.
Furthermore, if any particular value of $\phijlhat$ can be confidently ruled out for an event, then the source is inconsistent with having aligned spins.

This can have far-reaching consequences for the astrophysical interpretation of binary mergers.
The degree of spin--orbit alignment constrains the formation history of the system---in particular, whether it likely formed in isolation or in a dense environment \cite{Farr2018}---and can inform astrophysical processes such as tidal interactions and supernova kicks \cite{Gerosa2018}.
For this reason, characterizing the imprints of precession is a topic of intensive research.
Indeed, beyond the parametrization in terms of $\thetajn,\allowbreak \phi_{JL},\allowbreak \phi_{12}$ on which ours is built \cite{Farr2014}, numerous other descriptions have been studied, both physically and observationally motivated: in terms of an effective precession spin $\chi_{\rm p}$~\cite{Schmidt2015} or modifications thereof \cite{Gerosa2021, Thomas2021}; a precessing signal-to-noise ratio $\rho_{\rm p}$~\cite{Fairhurst2020}; a taxonomy of phenomenological parameters \cite{Gangardt2021}; or the spin azimuths \cite{Varma2021}.

The practical difference between $\phi_{JL}$ and $\phijlhat$ is that $\phi_{JL}$ posteriors typically have two modes separated by $\pi$, with opposite inclinations.
The reason for this bimodality is an approximate discrete symmetry corresponding to observing the same source from the antipodal direction, as discussed in Sec.~\ref{ssec:symmetries}.
As a result, the marginal posterior on $\phi_{JL}$ is not particularly well constrained.
We solve this by applying a shift of $\pi$ depending on the sign of $\cos\thetajn$ per Eq.~\eqref{eq:phijl_hat}.

Recently, \citet{Varma2021} found hints of precession in GWTC-2 by studying the azimuths $\phi_1, \phi_2$ about $\bm L$ between each individual spin and the orbital separation vector, at a reference time $t_{\rm ref}=-100\, GM/c^3$ before merger.
The results in Fig.~\ref{fig:spin_azimuth} are similar in spirit because, following an argument analogous to the one opening this section, a tight measurement of $\phi_1$ or $\phi_2$ would also rule out aligned spins.
In fact, we find that significantly tighter constraints can be placed on $\phijlhat$ than on $\phi_1$ or $\phi_2$.
This is for two reasons: first, the aforementioned symmetry leads to a similar bimodality correlated with the source inclination; second, $\phi_1$ and $\phi_2$ define the spin azimuths relative to the poorly measured orbital phase $\phiref$.
Indeed, \citet{Varma2021} report that varying the reference time (or equivalently, frequency) at which the source configuration is specified sensitively affects the constraints on $\phi_1, \phi_2$.
Interestingly, we find that the width of the $\phiref$ posterior depends on $\fref$, but that of $\phijlhat$ does not, at least in the range $\SI{50}{\hertz} < \fref < \SI{150}{\hertz}$.
This indicates that, at least in some cases, the uncertainty in $\phiref$ is the limiting factor in the measurement precision for the angle between spin and orbital separation.
We interpret this observation as follows.
First, while $\phi_1, \phi_2$ evolve at the orbital timescale, the angle $\phijlhat$ evolves at the slower precession timescale.
Moreover, the precession frequency is well constrained independently of directly observable precession effects, because intrinsic parameters are measured from the evolution of the orbital phase (note that to lowest post-Newtonian order, the precession frequency just depends on the masses and the orbital frequency, not the spins~\cite{Farr2014}).
Since there are few precession cycles and their frequency is determined, their phase can be specified through $\phijlhat$ similarly well over a broad range of reference frequencies.
This said, we observe a degradation in its measurement if we adopt a reference frequency of \SI{20}{\hertz}.
The azimuth difference $\phi_{12}$ also evolves on the precession timescale, however it has a subtler effect on the waveform and, in agreement with previous work \cite{Biscoveanu2021, Varma2021}, we find it is not as well-measured.

Reassuringly, the set of events with best measured $\phijlhat$ contains those for which precession signatures have been reported before, namely GW200129\_065458~\cite{Abbott2021_GWTC-3, Hannam2021}, GW190412 \cite{Abbott2020_GW190412, Hoy2021, Islam2021}, GW151226 \cite{Chia2022}, GW170818 \cite{Varma2021} and GW190521 \cite{Abbott2020_GW190521, Olsen2021}, as shown in Fig.~\ref{fig:spin_azimuth}.
That being said, at this point we are unable to quantify the significance of the measurability of $\phijlhat$ as an indicator of precession.
This preliminary result motivates future work, in which we will define a quantitative statistic involving $\phijlhat$ and calibrate its significance using synthetic data.

\section{Conclusions}
\label{sec:conclusions}

We have introduced a coordinate system optimized for characterization of compact binary mergers observed through gravitational waves.
It removes commonly encountered degeneracies and multimodality, and the transformation to standard coordinates has a simple Jacobian determinant and an explicit inverse.
These coordinates improve the robustness and efficiency of parameter estimation algorithms, and build intuition about gravitational wave measurements.

In order to remove degeneracy, the coordinates are designed to separately control the main observable features of the signal.
For the extrinsic parameters, these are the amplitude, phase and time of arrival at the reference (loudest) detector, and the time delay to the second-loudest detector.
For the spins, we single out the effective spin parameter and the total spin azimuth, which affect the orbital evolution and the peak frequencies of the amplitude modulations induced by precession, respectively.
We reemphasize previous realizations that the reference frequency at which the configuration of the binary is specified should be inside the sensitive band of the detectors, and the spin orientations should be defined relative to the direction of wave propagation rather than the orbital separation.

Strikingly, the azimuth $\phijlhat$ between the total spin and the unsigned direction of wave propagation is well measured in several examples (Fig.~\ref{fig:spin_azimuth}), establishing a novel observable signature of precession.
We anticipate that this parametrization will improve our understanding of spin--orbit misalignment in nature.
In future work we will define a statistic based on this parameter to quantify the significance of spin misalignment.

We identified four approximate symmetries as the leading cause of multimodality in posterior distributions, involving the orbital phase, polarization, sky location and inclination.
Depending on the signal-to-noise ratio and the network configuration, these can lead to up to $2^4$ modes in the posterior.
The last two of these symmetries are particularly good for the Hanford--Livingston network due to its peculiar geometric configuration, and can lead to four degenerate solutions roughly corresponding to a source overhead or underfoot, and face-on or face-off.

We devised ``folding'', an algorithm that turns a distribution with this type of multimodality into an equivalent unimodal problem by marginalizing over the discrete degeneracies.
To facilitate its application, we adapted our coordinates so that each of the four independent approximate-symmetry transformations is achieved by a one-parameter shift or reflection.

Using these algorithms, we were able to achieve robust parameter inference while keeping the number of live points, and thereby the computational cost, low (Fig.~\ref{fig:performance}).
We make publicly available a parameter estimation code that implements these features at \url{https://github.com/jroulet/cogwheel}.

These methods greatly simplify the distribution, to the point where a parametric approximation to the full distribution can be made based on the signal-to-noise ratio and a few inputs specifying the location of the peak.
Beyond the applications shown here, we expect that this will have other uses in gravitational wave data analysis.
For example, evidence integrals have application in search \cite{Smith2018,Olsen2022} and parameter estimation \cite{Pankow2015}, and knowledge of the distribution can be used to design efficient integration schemes, e.g.\ through variance reduction methods.
As another example, a machine-learning approach based on normalizing flows was recently demonstrated to be accurate and fast at estimating gravitational wave source parameters \cite{Dax2021}.
In that approach, the posterior distribution is described in terms of a system of coordinates in which it has a standard form (e.g.\ Gaussian).
This change of coordinates is found automatically and contains the complexity of the problem.
The techniques introduced in this paper can be regarded as an analytical approximation to a normalizing flow.
This suggests that applying a normalizing flow on the space of coordinates we developed here (or similar) might require a simpler neural network architecture and reduce the training cost.

\section*{Acknowledgements}

We thank Geraint Pratten for help with waveform models and Will Farr for comments on the manuscript.

This research has made use of data or software obtained from the Gravitational Wave Open Science Center (\url{gw-openscience.org}), a service of LIGO Laboratory, the LIGO Scientific Collaboration, the Virgo Collaboration, and KAGRA.
LIGO Laboratory and Advanced LIGO are funded by the United States National Science Foundation (NSF) as well as the Science and Technology Facilities Council (STFC) of the United Kingdom, the Max-Planck-Society (MPS), and the State of Niedersachsen/Germany for support of the construction of Advanced LIGO and construction and operation of the GEO600 detector.
Additional support for Advanced LIGO was provided by the Australian Research Council.
Virgo is funded, through the European Gravitational Observatory (EGO), by the French Centre National de Recherche Scientifique (CNRS), the Italian Istituto Nazionale di Fisica Nucleare (INFN) and the Dutch Nikhef, with contributions by institutions from Belgium, Germany, Greece, Hungary, Ireland, Japan, Monaco, Poland, Portugal, Spain.
The construction and operation of KAGRA are funded by Ministry of Education, Culture, Sports, Science and Technology (MEXT), and Japan Society for the Promotion of Science (JSPS), National Research Foundation (NRF) and Ministry of Science and ICT (MSIT) in Korea, Academia Sinica (AS) and the Ministry of Science and Technology (MoST) in Taiwan.

JR is supported by grant No.\ 216179 to the KITP from the Simons Foundation.
SO acknowledges support as an NSF Graduate Research Fellow under Grant No.\ DGE-2039656.
Any opinions, findings, and conclusions or recommendations expressed in this material are those of the authors and do not necessarily reflect the views of the National Science Foundation.
TI is  supported in part by the Heising-Simons Foundation, the Simons Foundation, and National Science Foundation Grant No.\ NSF PHY-1748958 and Grant No.\ PHY-2110496.
TV acknowledges support by the National Science Foundation under Grant No.\ 2012086.
BZ is supported by a research grant from the Center for New Scientists at the Weizmann Institute of Science and a research grant from the Ruth and Herman Albert Scholarship Program for New Scientists.
MZ is supported by the Canadian Institute for Advanced Research (CIFAR) program on Gravity and the Extreme Universe and the Simons Foundation Modern Inflationary Cosmology initiative.

\appendix

\section{Approximate symmetries, continued}
\label{app:symmetries}

In this appendix we provide a more quantitative discussion  of the argument presented in Sec.~\ref{ssec:symmetries}, regarding how the measurement of the Hanford--Livingston phase difference constrains the source azimuth along the time-delay ring and the discrete degeneracies associated.

\begin{figure}
    \centering
    \includegraphics[width=\linewidth]{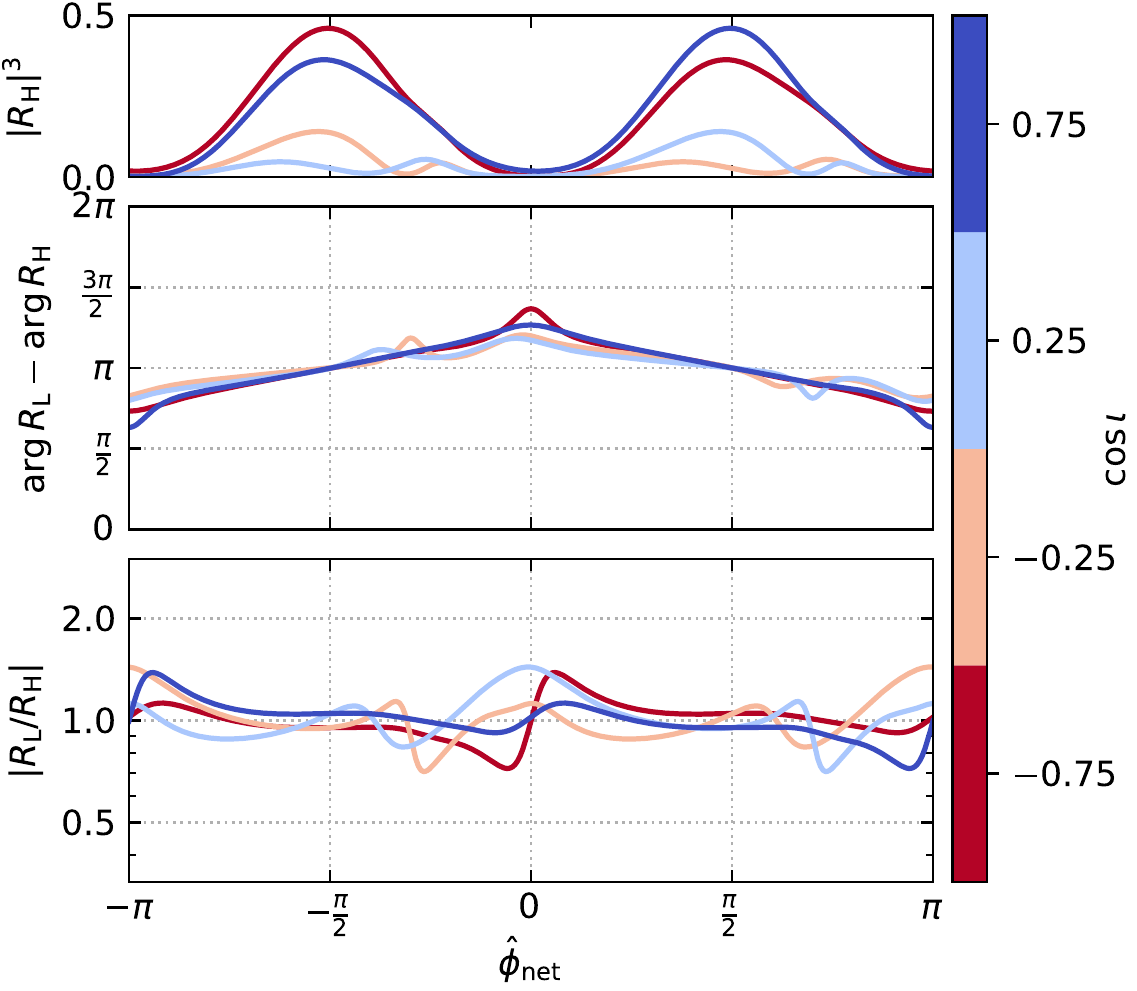}
    \caption{Response of the Hanford--Livingston network along a ring of constant time-delay, as a function of modified azimuth $\phinethat$ (Eq.~\eqref{eq:phinet_hat}) for various inclinations.\textit{ Top:} prior probability, see Eq.~\eqref{eq:dhat_prior}.\textit{ Center:} phase difference introduced by the relative orientations of the detectors.\textit{ Bottom:} relative amplitude response.
    In these plots, the transformation \eqref{eq:over_under_v2} is a horizontal reflection and \eqref{eq:faceon_faceoff_v2} inverts the color scale.
    Both are approximate symmetries of all three quantities.
    }
    \label{fig:azimuth_symmetry}
\end{figure}

From Eqs.~\eqref{eq:varphi_k} and \eqref{eq:dt_detectors}, the observable phase difference between detectors $k_0$ and $k_1$ is
\begin{equation}
\label{eq:phase_difference}
\begin{split}
    \varphi_{k_1} - \varphi_{k_0}
        &= \arg R_{k_1} - \arg R_{k_0}
        - 2 \pi \fbar_{k_1} \tau_{k_0 k_1} \cos\thetanet\\
        &\quad- 2 \pi (\fbar_{k_0} - \fbar_{k_1}) \trefdet,
\end{split}
\end{equation}
where the last term can be neglected if the detectors have similar noise power spectrum shapes and thus $\fbar_k$.
The difference $\arg R_{k_1} - \arg R_{k_0}$ does depend on the sky location and inclination if the detectors are not coaligned, which provides a joint constraint on $\thetanet, \phinet, \iota$.
Likewise, the relative amplitude at the detectors
\begin{equation}
    \frac{a_{k_1}}{a_{k_0}} = \abs{\frac{R_{k_1}}{R_{k_0}}}
\end{equation}
is also measurable.
Figure~\ref{fig:azimuth_symmetry} shows how these terms depend on $\phinethat$ for various inclinations (holding fixed the values of our other coordinates, and for an aligned-spin configuration so $\iota = \thetajn$).
Inverting $\cos\iota$ at fixed $\phinet$ inverts the sign of $\arg R_k$ and adds $\pi$ to $\phinethat$, by Eqs.~\eqref{eq:response} and \eqref{eq:phinet_hat}. Therefore, in the second panel blue and red curves are related by a vertical reflection plus a horizontal shift.
As a result of the geometry of the Hanford--Livingston network, the relative phase and amplitude of the detector responses are symmetric under the transformations \eqref{eq:symmetries_v2} to a good approximation, especially near the configurations favored by the prior $\phinethat \approx \pm \pi/2,\allowbreak \cos\iota \approx \pm 1$.

The dependence of $\arg R_{\rm L} - \arg R_{\rm H}$ on $\phinethat$ seen in Fig.~\ref{fig:azimuth_symmetry} is responsible for the correlation between 
$\cos\thetanet$ and $\phinethat$ noticeable in Fig.~\ref{fig:before_after}: per Eq.~\eqref{eq:phase_difference}, in order to match the observed $\varphi_{k_1} - \varphi_{k_0}$, $\cos\thetanet$ needs to change with $\phinethat$ to compensate the variation of $\arg R_{\rm L} - \arg R_{\rm H}$.

\section{Reference sheet}
\label{app:reference}

In this appendix we provide a compact list of the coordinates defined throughout the main text, as implemented in the \texttt{cogwheel} package.
The sequence of transformations to a standard system is given in Table~\ref{tab:coordinates}, together with their Jacobian determinants.
Here, GMST refers to the Greenwich Mean Sidereal Time of the event, which determines the orientation of Earth, and $k_0, k_1$ index the two detectors with highest signal-to-noise ratio.

\begin{table*}
    \setlength{\tabcolsep}{4pt}
    \centering
    \begin{tabular}{llllcc}
        \tabline
        Sampled & Standard  & \multicolumn{2}{c}{Conditioned on:} & $\abs{J}$ & Reference \\
        \cmidrule{3-4}
         &  &   \multicolumn{1}{c}{Variables} & \multicolumn{1}{c}{Constants} & \\
        \tabline
        $\mchirp, \ln q$    & $m_1, m_2$ & & & $\mchirp \cosh^{2/5}\left(\frac12\ln q\right)$\\
        $\chieff, \cumchidiff$ & $\sonez, \stwoz$  & $q$ & & \footnote{Unnecessary if the prior is chosen in terms of sampled parameters, e.g.\ uniform.} & \eqref{eq:chieff}, \eqref{eq:cumchidiff}\\
        $\cos\thetajn, \phijlhat, \phi_{12}, \csoner, \cstwor$ & $\iota, \sonexn, \soneyn, \stwoxn, \stwoyn$  & $\sonez, \stwoz, m_1, m_2$  &  $\fref$ & \footnote{Unnecessary since the prior is uniform on sampled parameters by isotropy and the definition of $\csoner, \cstwor$.} & \cite{Farr2014}, \eqref{eq:phijl_hat}, \eqref{eq:csoner}\\
        $\psi$   & $\psi$ &  &  & 1\\
        $\cos\thetanet, \phinethat$ & $\alpha, \sin\delta$  & $\thetajn$   & $\gmst, k_0, k_1$ & 1 & Fig.~\ref{fig:ligo_angles}, \eqref{eq:phinet_hat}\\
        $\trefdet$  & $\tc$ & $\alpha, \delta$  & $\gmst, k_0$ & 1 & \eqref{eq:t_k}\\
        $\phirefhat$   & $\phiref$ &  $\tc, \alpha, \delta, \psi, \iota$ & $\gmst, k_0, \fbar_{k_0}^{\rm ML}, \varphi_{k_0}^{\rm ML}$ & 1 & \eqref{eq:phirefhat_definition}\\
        $\dhat$ & $\dl$ & $\mchirp, \alpha, \delta, \psi, \iota$ & $\gmst, k_0$ & $\dhat / \dl$ & \eqref{eq:dhat} \\
        \tabline 
    \end{tabular}
    \caption{Modular sequence of transformations from the sampling coordinates we propose to a standard system.
    Each transformation involves few variables, is only conditioned on variables computed by the previous ones, and has a simple Jacobian determinant.}
    \label{tab:coordinates}
\end{table*}

We use $\sonexn, \soneyn, \stwoxn, \stwoyn$ as ``standard'' parameters to describe the in-plane spins; we define these as the Cartesian spins in a frame where $\bm{\hat z} \parallel \bm L$ and $\bm \propagation$ is in the $yz$-plane.
This is related to the ``radiation frame'' used by the LIGO Algorithm Library \cite{lalsuite} (where $\bm{\hat z} \parallel \bm L$ and $\bm{\hat x}$ is parallel to the orbital separation vector) with a rotation by $\phiref$ around $\bm {\hat z}$:
\begin{equation}
    \begin{pmatrix}
        \sonex & \stwox \\
        \soney & \stwoy
    \end{pmatrix} = \begin{pmatrix}
        \cos\phiref & \sin\phiref \\
        -\sin\phiref & \cos\phiref
    \end{pmatrix} \begin{pmatrix}
        \sonexn & \stwoxn \\
        \soneyn & \stwoyn       
    \end{pmatrix}.
\end{equation}
Using this system has two advantageous properties: the transformations in Table~\ref{tab:coordinates} get more decoupled since these spins are independent of $\phiref$, and the coprecessing-frame harmonic modes of a waveform transform under a change by $\phiref$ as $h_{\ell m}(f) \to h_{\ell m}e^{i m \phiref}$ if these spin components are held constant \cite{Bohe2016, Pratten2021} which is useful for reusing waveform computations, e.g.\ in folding.

We also introduced spin coordinates $\cumchidiff, \csoner, \cstwor$ which are the cumulatives of the prior on the aligned spin difference and the in-plane spin magnitudes, respectively.
These have a uniform prior on $(0, 1)$ by definition, and their relation to the physical spins depends on the choice of prior.
Our default choice is uniform in the aligned spin difference conditional on $\chieff, q$, in which case
\begin{equation}
    \begin{split}
    \label{eq:cumchidiff}
    \cumchidiff(\sonez, \stwoz, q)
    &\coloneqq \int_{\chi_{1z}^{\rm min}}^{\chi_{1z}} \rmd \sonez' \pi(\sonez' \mid \chieff, q) \\
    &= \frac{\chi_{1z} - \chi_{1z}^{\rm min}}{\chi_{1z}^{\rm max} - \chi_{1z}^{\rm min}},
    \end{split}
\end{equation}
where $\chi_{1z}^{\rm min}(\sonez, \stwoz, q)$ is the minimum possible $\sonez$ consistent with the value of effective spin $\chieff(\sonez, \stwoz, q)$ (given by Eq.~\eqref{eq:chieff}) subject to the Kerr bound $\abs{\sonez} < 1$, and similarly $\chi_{1z}^{\rm max}(\sonez, \stwoz, q)$ is the maximum possible value:
\begin{equation}
    \begin{split}
        \chi_{1z}^{\rm min} &= \max(\sonez + q \, \stwoz - q, -1) \\
        \chi_{1z}^{\rm max} &= \min(\sonez + q \, \stwoz + q, 1).
    \end{split}
\end{equation}
For the in-plane spins, our prior choice is uniform in the disk given the aligned spin value, which yields
\begin{equation}
    \begin{split}
        \label{eq:csoner}
        \csoner(\soner, \sonez) &\coloneqq \int_0^{\soner} \rmd {\soner}'\pi({\soner}' \mid \sonez) \\
        &= \frac{(\soner)^2}{1 - \sonez^2},
    \end{split}
\end{equation}
and similarly for the secondary in-plane spin.

\bibliography{main}

\end{document}